\documentclass[13 pt]{article}
\usepackage{amssymb}

\def\hcal{{\Large{\textit{h}}}}
\def\dcal{{\Large{\textit{d}}}}
\def\acal{{\Large{\textit{a}}}}
\def\bcal{{\Large{\textit{b}}}}
\def\ccal{{\Large{\textit{c}}}}
\def\xcal{{\Large{\textit{x}}}}
\def\acalt{\tilde{\Large{\textit{a}}}}
\def\bcalt{\tilde{\Large{\textit{b}}}}
\def\ccalt{\tilde{\Large{{\textit{c}}}}}
\def\Acal{{{{\cal A}}}}
\def\Bcal{{{{\cal B}}}}
\def\Ccal{{{{\cal C}}}}
\def\Xcal{{{{\cal X}}}}
\def\Acalt{\tilde{{{\cal A}}}}
\def\Bcalt{\tilde{{{\cal B}}}}
\def\Ccalt{\tilde{{{{\cal C}}}}}
\def\Xcalt{\tilde{{{{\cal X}}}}}
\def\dcalt{\tilde{\Large{{\textit{d}}}}}
\def\hcalt{\tilde{\Large{{\textit{h}}}}}
\def\xcalt{\tilde{\Large{{\textit{x}}}}}

\def\binombN{\left(^{N}_{n+1}\right)}
\def\binomaN{\left(^{N}_{n}\right)}
\def\binomcN{\left(^{N}_{n-1}\right)}
\def\binomNmun{\left(^{N-1}_{n}\right)}
\def\binomN{\left(^{N}_{n}\right)}
\def\binomNmunpun{\left(^{N-1}_{n+1}\right)}
\def\binomNpun{\left(^{N}_{n+1}\right)}
\def\binomNmunmun{\left(^{N-1}_{n-1}\right)}
\def\binomNnmun{\left(^{N}_{n-1}\right)}

\def\binomNs{\left(^{N}_{s}\right)}
\def\binomNspun{\left(^{N}_{s+1}\right)}
\def\binomNsmun{\left(^{N}_{s-1}\right)}

\def\lambdat{\tilde{\lambda}}

\newcommand{\beq}{\begin{equation}}
\newcommand{\eeq}{\end{equation}}
\newcommand{\bea}{\begin{eqnarray*}}
\newcommand{\eea}{\end{eqnarray*}}
\newcommand{\beqa}{\begin{eqnarray}}
\newcommand{\eeqa}{\end{eqnarray}}

\begin{document}

\newfont{\elevenmib}{cmmib10 scaled\magstep1}%

\newcommand{\Title}[1]{{\baselineskip=26pt \begin{center}
            \Large   \bf #1 \\ \ \\ \end{center}}}
\hspace*{2.13cm}%
\hspace*{1cm}%
\newcommand{\Author}{\begin{center}\large
           Pascal Baseilhac\footnote{
baseilha@phys.univ-tours.fr} and Kozo Koizumi\footnote{kkoizumi@cc.kyoto-su.ac.jp}
\end{center}}
\newcommand{\Address}{{\baselineskip=18pt \begin{center}
$^1$\it Laboratoire de Math\'ematiques et Physique Th\'eorique CNRS/UMR 6083,\\
F\'ed\'eration Denis Poisson,\\
Universit\'e de Tours, Parc de Grandmont, 37200 Tours, France\vspace{0.4mm}\\

$^2$\it Department of Physics,\\
Kyoto Sangyo University, Kyoto 603-8555, Japan
\end{center}}}
\baselineskip=13pt

\bigskip
\vspace{-1cm}

\Title{Exact spectrum of the XXZ open spin chain from\\ the $q-$Onsager algebra representation theory}\Author

\vspace{- 0.1mm}
 \Address

\vskip 0.6cm

\centerline{\bf Abstract}\vspace{0.3mm}  \vspace{1mm}
The transfer matrix of the XXZ open spin-$\frac{1}{2}$ chain with general integrable boundary conditions and generic anisotropy parameter ($q$ is not a root of unity and $|q|=1$) is diagonalized using the representation theory of the $q-$Onsager algebra. Similarly to the Ising and superintegrable chiral Potts models, the complete spectrum is expressed in terms of the roots of a characteristic polynomial of degree $d=2^N$. The complete family of eigenstates are derived in terms of rational functions defined on a discrete support which satisfy a system of coupled recurrence relations. In the special case of linear relations between left and right boundary parameters for which Bethe-type solutions are known to exist, our analysis provides an alternative derivation of the results of Nepomechie {\it et al.} and Cao {\it et al.}. 
In the latter case the complete family of eigenvalues and eigenstates splits in two sets, each associated with a characteristic polynomial of degree $d< 2^N$. Numerical checks performed for small values of $N$ support the analysis.   
\vspace{0.1cm} 

\vskip 0.8cm

\vskip -0.6cm


\section{Introduction}
Since the pioneering work by Bethe on the Heisenberg isotropic chain \cite{Bethe}, various methods and techniques have been proposed to solve quantum integrable models on the lattice. Notably, all of them take their roots in Bethe's article or Onsager's articles on the planar Ising model \cite{Ons,Kauf}. By now standard approaches, the so-called algebraic and functional Bethe ansatz have been successfully applied to a large number of quantum integrable models defined through a transfer matrix or Hamiltonian, for instance quantum spin chains or vertex models with periodic boundary conditions. In both approaches, the transfer matrix eigenvalues are written in terms of solutions of Bethe equations  whereas the construction of the so-called Bethe eigenstates requires either the existence of a simple reference state (also called pseudo-vacuum state) on which the creation operator acts iteratively, or the construction of modified Baxter$-Q$ operators. In any case, the proof of completeness of the eigenstates is not straightforward and makes necessary numerical computations.

The homogeneous XXZ open spin-$\frac{1}{2}$ chain with the most general integrable boundary conditions is among the models for which these standard techniques cannot be applied, except for some special family of boundary conditions or special values of the anisotropy parameter. Its Hamiltonian reads
\beqa
H_{XXZ}&=&\sum_{k=1}^{N-1}\Big(\sigma_1^{k}\sigma_1^{k+1}+\sigma_2^{k}\sigma_2^{k+1} + \Delta\sigma_3^{k}\sigma_3^{k+1}\Big) \nonumber\\
&&+\ \frac{(q^{1/2}-q^{-1/2})}{(\epsilon^{(0)}_+ + \epsilon^{(0)}_-)}
\Big(  \frac{(\epsilon^{(0)}_+ - \epsilon^{(0)}_-)}{2}\sigma^1_3 + \frac{2}{(q^{1/2}-q^{-1/2})}\big(k_+\sigma^1_+ + k_-\sigma^1_-\big)       \Big)\nonumber\\
 &&+\ \frac{(q^{1/2}-q^{-1/2})}{(\kappa + \kappa^*)}
\Big(  \frac{(\kappa - \kappa^*)}{2}\sigma^N_3 + 2(q^{1/2}+q^{-1/2})\big(\kappa_+\sigma^N_+ + \kappa_-\sigma^N_-\big)       \Big)\label{H}\ ,
\eeqa
where $\sigma_{1,2,3}$ and $\sigma_\pm=(\sigma_1\pm i\sigma_2)/2$ are usual Pauli matrices and $\Delta=(q^{1/2}+q^{-1/2})/2$\ \ denotes the anisotropy parameter.  In the present analysis, we will restrict our attention to the (massless) regime $-1\leq\Delta\leq 1$ i.e. $q=\exp(\phi)$ with $\phi$ purely imaginary. To exhibit the {\it six} independent boundary parameters, below we will sometimes use the following parametrization with $\theta,{\tilde\theta}\in{\mathbb R}$ and ${\alpha}, {\tilde\alpha}\in{\mathbb C}$:
\beqa
&&\epsilon^{(0)}_{+}=(\epsilon^{(0)}_{-})^{\dagger}=\cosh\alpha\ ,\qquad \qquad k_+=(k_{-})^{\dagger}=-(q^{1/2}-q^{-1/2})e^{i\theta}/2 \qquad\ \quad  \ \mbox{(left)}\ ,\nonumber\\ 
&&\kappa^*=(\kappa)^{\dagger}=-\cosh{\tilde\alpha}\ ,\quad \ \qquad \ \ \
\kappa_+=-(\kappa_{-})^{\dagger}=-e^{i{\tilde{\theta}}}/(2(q^{1/2}+q^{-1/2}))\qquad \mbox{(right)}\ .\label{param}
\eeqa

For the special case of diagonal boundary conditions considered in \cite{Sklya,Alca}, or for special linear relations between the left and right boundary parameters \cite{Nepo,Cao} (see also \cite{WL}) this model has been studied within the algebraic Bethe ansatz or functional Bethe ansatz frameworks. More recently, a Bethe-type solution has also been obtained for arbitrary (generic) boundary parameters provided the anisotropy parameter takes special values ($q$ is a root of unity) \cite{Nepo2}.
However, for generic integrable boundary conditions and anisotropy parameter the exact spectrum of (\ref{H}) has remained an outstanding problem. Either due to the lack of a pseudo-vacuum state, or due to the lack of functional relations of finite order for the transfer matrix, both algebraic and functional Bethe ansatz techniques have failed in this regime of parameters.

In this paper, a solution to this problem is given: we propose the exact solution of the model (\ref{H}) for arbitrary parameters and $q$ not a root of unity.
The approach here presented, initiated in \cite{qDG}, partly takes its roots in Onsager's original work \cite{Ons}. Mainly based on the representation theory of the infinite dimensional $q-$Onsager algebra (also called tridiagonal algebra) which underlies the integrability of the model \cite{qOns}, our results for the spectrum and eigenstates essentially differ in form from the ones usually obtained using Bethe ansatz techniques. As expected from the method used, they have closed analogy with the ones for the planar Ising and superintegrable chiral Potts models \cite{Potts,Davies}.   

In the next Section, we recall how the transfer matrix of the XXZ open spin chain (\ref{H}) can be simply written in terms of the generators of the $q-$deformed analogue of the Onsager algebra and its Abelian subalgebra. Then, these generators are shown to possess a block tridiagonal structure in the vector and functional basis proposed in \cite{TDpair} for which all coefficients are derived recursively (see Appendix A). In Section 3,   
the spectral problem for the transfer matrix is then considered in details: the eigenstates - written in the vector basis - are derived as linear combinations of rational functions of one discrete argument which takes its value within the roots of a characteristic polynomial of degree $d=2^N$. Then, we consider the special case of linear relations among the boundary parameters and discuss the correspondence with each of the two complementary sets of Bethe-type solutions obtained in \cite{Nepo,Cao,WL} which, all together, cover the full spectrum. For general integrable boundary conditions, all the eigenvalues of the transfer matrix are derived from second-order discrete $q-$difference equations.  Concluding remarks follow in the last Section. In Appendix B, numerical results for the case of $N=3$ sites are reported as an example.

\section{The transfer matrix in $q-$deformed Onsager's basis}
\subsection{An alternative description}
In Sklyanin's approach of boundary quantum integrable models, the transfer matrix of the XXZ open spin chain is constructed starting from a $R-$matrix - which statisfies the Yang-Baxter equation - and two $K_{\pm}-$matrices - which satisfy the reflection and dual reflection equations, respectively - associated with the two boundaries. In particular, from the most general $c-$number $K-$matrices derived in \cite{DeVeg} one obtains the Hamiltonian (\ref{H}) among the {\it local} conserved quantities. However, as pointed out in \cite{qDG} the transfer matrix can be alternatively written in terms of the generators of an Abelian subalgebra of the $q-$deformed analogue of the Onsager algebra (see details below) with no necessary reference to Yang-Baxter-type of algebraic structures. In this point of view, it reads   
\beqa
t_{XXZ}(u)= \sum_{l=0}^{N-1}{\cal F}_{2l+1}(u)\ {{\cal I}}_{2l+1}^{(N)} + {\cal F}_0(u)\ I\!\!I^{(N)}\label{tfin}
\eeqa
where
\beqa
{\cal F}_{2l+1}(u)&=&-\frac{1}{(q^{1/2}+q^{-1/2})}\frac{(qu^4+q^{-1}u^{-4}-q-q^{-1})}{(q+q^{-1}-u^2-u^{-2})^N}
\sum_{n=l}^{N-1}\Big(\frac{q^{1/2}u^2+q^{-1/2}u^{-2}}{q^{1/2}+q^{-1/2}}\Big)^{n-l}C_{-n}^{(N)}            \ ,\nonumber\\
{\cal F}_{0}(u)&=&\frac{(qu^4+q^{-1}u^{-4}-q-q^{-1})(q^{1/2}+q^{-1/2})}{(q+q^{-1}-u^2-u^{-2})^N}\ \nonumber \\
&&\ \qquad\times \  \frac{k_+k_-}{(q^{1/2}-q^{-1/2})}\Big(\frac{\kappa_+}{k_+}+\frac{\kappa_-}{k_-}\Big) \left(\frac{{\omega}_0^{(N)}(q^{1/2}-q^{-1/2})}{k_+k_-}-\sum_{n=0}^{N-1}\Big(\frac{q^{1/2}u^2+q^{-1/2}u^{-2}}{q^{1/2}+q^{-1/2}}\Big)^{n+1}C_{-n}^{(N)} \right)\nonumber\\
&&\ \ + \ \frac{\big((q^{1/2}u^2+q^{-1/2}u^{-2})\kappa+ (q^{1/2}+q^{-1/2})\kappa^*\big)\epsilon_+^{(N)}}{(q+q^{-1}-u^2-u^{-2})^N}\ \nonumber\\
&&\ \ + \ \frac{\big((q^{1/2}u^2+q^{-1/2}u^{-2})\kappa^*+ (q^{1/2}+q^{-1/2})\kappa\big)\epsilon_-^{(N)}}{(q+q^{-1}-u^2-u^{-2})^N}\ \nonumber
\eeqa
is a family of rational functions in the spectral parameter $u$. The explicit expressions for the coefficients $C_{-n}^{(N)}$ and $\omega_0^{(N)}$, $\epsilon^{(N)}_{\pm}$ essentially depend on the choice of quantum space representations at each site (two-dimensional for (\ref{H})) of the spin chain. Calculated recursively in \cite{qOns} (see also \cite{qOns0}), for the model (\ref{H}) they read 
\beqa
C_{-n}^{(N)}=(-1)^{N-n} (q^{1/2}+q^{-1/2})^{n+1}\sum_{k_1<...<k_{N-n-1}=1}^{N}\!\!\!\!\!\alpha_{k_1}\cdot \cdot \cdot\alpha_{k_{N-n-1}}\label{coefflin}
\eeqa
where we have defined
\beqa
\alpha_1&=& \frac{2(q+q^{-1})}{(q^{1/2}+q^{-1/2})}+ \frac{\epsilon^{(0)}_{+}\epsilon^{(0)}_{-}(q^{1/2}-q^{-1/2})^2}{k_+k_-(q^{1/2}+q^{-1/2})}\ ,\qquad \alpha_{k}=\frac{2(q+q^{-1})}{(q^{1/2}+q^{-1/2})}\quad \mbox{for}\quad k=2,...,N\ ,\nonumber
\eeqa
%
%
%
%
and for arbitrary values of $N$ 
\beqa
\epsilon^{(N)}_{\pm}&=&(q+q^{-1})\epsilon^{(N-1)}_{\mp} - 2\epsilon^{(N-1)}_{\pm}\ ,\nonumber\\
\omega_0^{(N)}&=&(-1)^{N}\frac{k_+k_-}{q^{1/2}-q^{-1/2}}\Big(\frac{2(q+q^{-1})}{(q^{1/2}+q^{-1/2})}\Big)^{N-1} \left( \frac{2(q+q^{-1})}{(q^{1/2}+q^{-1/2})}+ \frac{\epsilon^{(0)}_{+}\epsilon^{(0)}_{-}(q^{1/2}-q^{-1/2})^2}{k_+k_-(q^{1/2}+q^{-1/2})}\right)\ .\label{constdef} \eeqa
\vspace{2mm}

On the other hand, the algebraic part of (\ref{tfin}) for the model (\ref{H}) is formed by a family of mutually commuting operators acting on the tensor product of $N$ two-dimensional quantum spaces. Generalizing the Onsager's hierarchy, they form an Abelian subalgebra \cite{qOns}. Indeed, similarly to the undeformed case each operator ${{\cal I}}_{2l+1}^{(N)}$  can be written in terms of the generators of the $q-$deformed analogue of the Onsager algebra introduced and studied in \cite{qOns0}. Using the basis of $4N$ operators ${\cal W}^{(N)}_{-l} ,{\cal W}^{(N)}_{l+1}, {\cal G}^{(N)}_{l+1}, {\tilde{\cal G}}^{(N)}_{l+1}$ with $l=0,1,...,N-1$ introduced in \cite{qOns}, they take the simple form\,\footnote{Note that the remaining boundary parameters are implicitely contained in the realizations of ${\cal W}^{(N)}_{-l} ,{\cal W}^{(N)}_{l+1}, {\cal G}^{(N)}_{l+1}, {\tilde{\cal G}}^{(N)}_{l+1}$.}
\beqa
{{\cal I}}_{2l+1}^{(N)}=\kappa {\cal W}^{(N)}_{-l} + \kappa^* {\cal W}^{(N)}_{l+1} + \frac{\kappa_+}{k_+} {\cal G}^{(N)}_{l+1} 
+ \frac{\kappa_-}{k_-} {\tilde{\cal G}}^{(N)}_{l+1}\ .\label{Ifin}
\eeqa

The fact that all operators ${{\cal I}}_{2l+1}^{(N)}$, $l=0,1,...,N-1$ are in involution is actually a direct consequence of the defining relations (see details below) of the $q-$deformed analogue of the Onsager algebra \cite{qOns0}. In view of the expansion (\ref{tfin}) it follows
\beqa
\big[{\cal I}^{(N)}_{2k+1},{\cal I}^{(N)}_{2l+1}\big]=0  \qquad \Rightarrow     \qquad \big[H_{XXZ},{\cal I}^{(N)}_{2k+1}\big]=0  \qquad\qquad \mbox{for all} \qquad k,l\in 0,...,N-1\ \label{commut}  
\eeqa
which characterize all Abelian symmetries of the XXZ open spin chain (\ref{H}) \cite{qOns}. More generally and independently of the explicit form of the rational functions ${\cal F}_{2l+1}(u)$, from (\ref{commut}) and (\ref{tfin}) one easily recovers the well-known integrability condition for (\ref{H}): 
\beqa
\big[t_{XXZ}(u),t_{XXZ}(v)\big]=0\ .\label{inv}
\eeqa

According to (\ref{tfin}), the spectral problem i.e. finding the eigenstates and eigenvalues of the transfer matrix (as well as any {\it local} or {\it nonlocal} conserved quantities) reduces to the one for the family of mutually commuting operators (\ref{Ifin}). In view of (\ref{Ifin}), this problem clearly needs a detailed analysis of the finite dimensional representations of the $q-$deformed analogue of the Onsager algebra which is considered below.

\subsection{The $q-$deformed analogue of the Onsager algebra and tensor product representations}
In \cite{qOns0}, based on the structure of the Sklyanin operator a $q-$deformed analogue of the Onsager algebra has been introduced and studied in details. With elements ${\textsf W}_{-k}$, ${\textsf W}_{k+1}$, ${\textsf G}_{k+1}$, ${\tilde{\textsf G}}_{k+1}$, it is a non-Abelian infinite dimensional algebra with defining relations
\beqa
&&\big[{\textsf W}_0,{\textsf W}_{k+1}\big]=\big[{\textsf W}_{-k},{\textsf W}_{1}\big]=\frac{1}{(q^{1/2}+q^{-1/2})}\big({\tilde{\textsf G}_{k+1} } - {{\textsf G}_{k+1}}\big)\ ,\nonumber\\
&&\big[{\textsf W}_0,{\textsf G}_{k+1}\big]_q=\big[{\tilde{\textsf G}}_{k+1},{\textsf W}_{0}\big]_q=\rho{\textsf W}_{-k-1}-\rho{\textsf W}_{k+1}\ ,\nonumber\\
&&\big[{\textsf G}_{k+1},{\textsf W}_{1}\big]_q=\big[{\textsf W}_{1},{\tilde{\textsf G}}_{k+1}\big]_q=\rho{\textsf W}_{k+2}-\rho{\textsf W}_{-k}\ ,\nonumber\\
&&\big[{\textsf W}_0,{\textsf W}_{-k}\big]=0\ ,\quad 
\big[{\textsf W}_1,{\textsf W}_{k+1}\big]=0\ ,\quad \nonumber\\
&&\big[{\textsf G}_{k+1},{\textsf G}_{l+1}\big]=0\ ,\quad   \big[{\tilde{\textsf G}}_{k+1},\tilde{{\textsf G}}_{l+1}\big]=0\ ,\quad
\big[{\tilde{\textsf G}}_{k+1},{\textsf G}_{l+1}\big]
+\big[{{\textsf G}}_{k+1},\tilde{{\textsf G}}_{l+1}\big]=0\ \label{qOns}
\eeqa
and fixed scalar $\rho$, $k,l\in {\mathbb N}$, where the $q-$commutator \ $[X,Y]_{q}=q^{1/2}XY-q^{-1/2}YX$\ has been introduced. In particular, assuming\,\footnote{It is the case for the corresponding operators of the XXZ open spin chain (\ref{H}).} \ ${\textsf G}_1=[{\textsf W}_1,{{\textsf W}_0}]_q$ \ and\  ${\tilde{\textsf G}}_{1}=[{\textsf W}_0,{{\textsf W}_1}]_q$ \ it is not difficult to exhibit a close relationship between the algebra (\ref{qOns}) and the $q-$Onsager algebra\,\footnote{However, an isomorphism between the $q-$deformed analogue of the Onsager algebra (\ref{qOns}) and the $q-$Onsager/Tridiagonal algebra has not been exhibited yet.} (also called tridiagonal algebra introduced in \cite{Ter}). Indeed, replacing these elements in the fourth relations of (\ref{qOns}) for $k=1$ and using the second and third relations one immediately recognizes the so-called $q-$deformed Dolan-Grady relations discussed in \cite{Ter} (see also \cite{qDG}).\vspace{1mm}

For the XXZ open spin$-\frac{1}{2}$ chain with generic boundary conditions (\ref{H}), the algebra (\ref{qOns}) is a hidden dynamical (nonlocal) symmetry of the model: It has been shown that the operators  ${\cal W}^{(N)}_{-l} ,{\cal W}^{(N)}_{l+1}, {\cal G}^{(N)}_{l+1}, {\tilde{\cal G}}^{(N)}_{l+1}$ in (\ref{Ifin}) induce a representation of (\ref{qOns}) provided one identifies \cite{qOns}
\beqa
\rho = (q^{1/2}+q^{-1/2})^2 k_+k_-\ .\label{rho}
\eeqa\nonumber
By comparison with the case of the Onsager algebra, for finite dimensional representations linear relations among the realizations of the elements generating (\ref{qOns}) are well-expected.  Indeed, for the finite dimensional representations associated with the model (\ref{H}) we have found \cite{qOns}
\beqa
&&-\frac{(q^{1/2}-q^{-1/2})}{k_+k_-}\omega_0^{(N)}{\cal W}_{0}^{(N)}+\sum^{N}_{l=1}C_{-l+1}^{(N)}{\cal W}_{-l}^{(N)} + \epsilon^{(N)}_{+}I\!\!I^{(N)}=0\ ,\nonumber\\
&&-\frac{(q^{1/2}-q^{-1/2})}{k_+k_-}\omega_0^{(N)}{\cal W}_{1}^{(N)}+\sum^{N}_{l=1}C_{-l+1}^{(N)}{\cal W}_{l+1}^{(N)} + \epsilon^{(N)}_{-}I\!\!I^{(N)}=0\ ,\nonumber\\
&&-\frac{(q^{1/2}-q^{-1/2})}{k_+k_-}\omega_0^{(N)}{\cal G}_{1}^{(N)}+\sum^{N}_{l=1}C_{-l+1}^{(N)}{\cal G}_{l+1}^{(N)}=0\ ,\nonumber\\
&&-\frac{(q^{1/2}-q^{-1/2})}{k_+k_-}\omega_0^{(N)}{\tilde {\cal G}}_{1}^{(N)}+\sum^{N}_{l=1}C_{-l+1}^{(N)}{\tilde {\cal G}}_{l+1}^{(N)}=0\ \label{c4}\
\eeqa
with (\ref{coefflin}), (\ref{constdef}).\vspace{1mm} 

Let us finally describe explicitely the structure of the operators ${\cal W}^{(N)}_{-l} ,{\cal W}^{(N)}_{l+1}, {\cal G}^{(N)}_{l+1}, {\tilde{\cal G}}^{(N)}_{l+1}$ acting on the tensor product representation ${\cal V}: ({\mathbb C}^{2})^{\otimes N}$. From the results of \cite{qOns} which were based on    
the representation theory of the $U_{q^{1/2}}(sl_2)$ algebra and its connection with the representation theory of (\ref{qOns}), one has
\beqa
{\cal W}_{-l}^{(N)}&=&\frac{(q+q^{-1})}{(q^{1/2}+q^{-1/2})}I\!\!I\otimes
{\cal W}_{l}^{(N-1)}
-\frac{2}{(q^{1/2}+q^{-1/2})}I\!\!I\otimes {\cal W}_{-l+1}^{(N-1)}+\ q^{\sigma_3/2}\otimes \big({\cal W}_{-l}^{(N-1)}-{\cal W}_{l}^{(N-1)}\big)
\nonumber\\
&&\ \ \ + \ \frac{(q^{1/2}-q^{-1/2})}{k_+k_-(q^{1/2}+q^{-1/2})^2}
\left(k_+\sigma_+\otimes
{\cal G}_{l}^{(N-1)}+k_-\sigma_-\otimes {\tilde {\cal G}}_{l}^{(N-1)}\right)+ \frac{2(q+q^{-1})}{(q^{1/2}+q^{-1/2})^2}{\cal W}_{-l+1}^{(N)}\ ,\nonumber\\
{\cal W}_{l+1}^{(N)}&=&\frac{(q+q^{-1})}{(q^{1/2}+q^{-1/2})}I\!\!I\otimes
{\cal W}_{-l+1}^{(N-1)}
-\frac{2}{(q^{1/2}+q^{-1/2})}I\!\!I\otimes {\cal W}_{l}^{(N-1)}+\ q^{-\sigma_3/2}\otimes \big({\cal W}_{l+1}^{(N-1)}-{\cal W}_{-l+1}^{(N-1)}\big)
\nonumber\\
&&\ \ \ + \ \frac{(q^{1/2}-q^{-1/2})}{k_+k_-(q^{1/2}+q^{-1/2})^2}
\left(k_+\sigma_+\otimes
{\cal G}_{l}^{(N-1)}+k_-\sigma_-\otimes {\tilde {\cal G}}_{l}^{(N-1)}\right)+ \frac{2(q+q^{-1})}{(q^{1/2}+q^{-1/2})^2}{\cal W}_{l}^{(N)}\ ,\nonumber\\
{\cal G}_{l+1}^{(N)}&=&   I\!\!I\otimes \big({\cal G}_{l+1}^{(N-1)}-{\cal G}_{l}^{(N-1)}\big) + \frac{2(q+q^{-1})}{(q^{1/2}+q^{-1/2})^2}{\cal G}_{l}^{(N)}\ 
\nonumber\\
&& \quad + \ k_-(q-q^{-1})\sigma_-\otimes \big({\cal W}_{-l}^{(N-1)}+{\cal W}_{l+1}^{(N-1)}-{\cal W}_{l}^{(N-1)}-{\cal W}_{-l+1}^{(N-1)}\big)\ ,
\nonumber\\
{\tilde{\cal G}}_{l+1}^{(N)}&=& I\!\!I\otimes \big({\tilde{\cal G}}_{l+1}^{(N-1)}-{\tilde{\cal G}}_{l}^{(N-1)}\big) + \frac{2(q+q^{-1})}{(q^{1/2}+q^{-1/2})^2}{\tilde{\cal G}}_{l}^{(N)}\ 
\nonumber\\
&& \quad + \ k_+(q-q^{-1})\sigma_+\otimes \big({\cal W}_{-l}^{(N-1)}+{\cal W}_{l+1}^{(N-1)}-{\cal W}_{l}^{(N-1)}-{\cal W}_{-l+1}^{(N-1)}\big)\ .\label{rep}
\eeqa
For the special case $l=0$ note that we identify\,\footnote{The reader must keep in mind that ${{\cal W}}_{l}^{(N)}|_{l=0}\neq {{\cal W}}_{-l}^{(N)}|_{l=0}$\ ,${{\cal W}}_{-l+1}^{(N)}|_{l=0}\neq {{\cal W}}_{l+1}^{(N)}|_{l=0}$\ for any $N$.}
\beqa
{{\cal W}}_{l}^{(N)}|_{l=0}\equiv 0\ ,\quad {{\cal W}}_{-l+1}^{(N)}|_{l=0}\equiv 0\ ,\quad {\cal G}_{l}^{(N)}|_{l=0}={\tilde{\cal G}}_{l}^{(N)}|_{l=0}\equiv \frac{k_+k_-(q^{1/2}+q^{-1/2})^2}{(q^{1/2}-q^{-1/2})}I\!\!I^{(N)}\ \label{not}
\eeqa
and the identity operator $I\!\!I^{(N)}=I\!\!I\otimes \cdot \cdot \cdot \otimes I\!\!I$
has been introduced. In addition, one has the ``initial'' $c-$number conditions
\beqa
{{\cal W}}_{0}^{(0)}\equiv \epsilon^{(0)}_+\ ,\quad {{\cal W}}_{1}^{(0)}\equiv \epsilon^{(0)}_-\qquad  \mbox{and}\qquad
{\cal G}_{1}^{(0)}={\tilde{\cal G}}_{1}^{(0)}\equiv \epsilon^{(0)}_+\epsilon^{(0)}_-(q^{1/2}-q^{-1/2})\ .\label{initrep}
\eeqa

Together with (\ref{c4}), the expressions above determine the action of the nonlocal operators ${{\cal I}}_{2l+1}^{(N)}$ given by (\ref{Ifin}) on ${\cal V}$.
Being of a rather complicated form, below we use different basis introduced in \cite{TDpair} for ${\cal V}$ on which all nonlocal operators act in a rather simple manner. 

\subsection{Block tridiagonal structure of the $q-$Onsager generators and their descendents}
As shown in \cite{TDpair}, the family of nonlocal operators ${\cal W}_{0}^{(N)},{\cal W}_{1}^{(N)}$ enjoy remarkable properties. For the XXZ open spin chain with general boundary conditions (\ref{H}), using the definitions (\ref{rep}) it is actually not difficult to show that 
\beqa
{\cal G}^{(N)}_1=[{\cal W}^{(N)}_1,{{\cal W}^{(N)}_0}]_q\ ,\qquad {\tilde{\cal G}}^{(N)}_{1}=[{\cal W}^{(N)}_0,{{\cal W}_1}^{(N)}]_q\ .\label{G1}
\eeqa
Consequently, using (\ref{qOns}) one finds that ${\cal W}_{0}^{(N)},{\cal W}_{1}^{(N)}$ also induce a representation of the {\it $q-$Onsager} i.e. {\it tridiagonal algebra} \cite{Ter} with  defining relations (so-called $q-$deformed Dolan-Grady relations) and (\ref{rho}) as:
\beqa \big[{\cal W}^{(N)}_0,\big[{\cal W}^{(N)}_0,\big[{\cal W}^{(N)}_0,{\cal
W}^{(N)}_1\big]_q\big]_{q^{-1}}\big]&=&\rho\big[{\cal W}^{(N)}_0,{\cal W}^{(N)}_1\big]\ ,\nonumber\\
\big[{\cal W}^{(N)}_1,\big[{\cal W}^{(N)}_1,\big[{\cal W}^{(N)}_1,{\cal
W}^{(N)}_0\big]_q\big]_{q^{-1}}\big]&=&\rho\big[{\cal W}^{(N)}_1,{\cal W}^{(N)}_0\big]\ .
\label{Talg}
\eeqa
In other words, the pair of operators ${\cal W}_{0}^{(N)},{\cal W}_{1}^{(N)}$ is said to be a {\it tridiagonal pair} which general definition can be found in \cite{Ter}. Briefly, for generic boundary parameters (\ref{param}) and generic  values of $q$ there exists a complete basis (resp. dual basis) of ${\cal V}$ in which  ${\cal W}_{0}^{(N)}$ (resp. ${\cal W}_{1}^{(N)}$) is represented by a diagonal matrix with degeneracies and ${\cal W}_{1}^{(N)}$ (resp. ${\cal W}_{0}^{(N)}$) is represented by a block tridiagonal matrix. Such basis and dual one have been constructed in \cite{TDpair}, so that we refer the reader to this work for details and only recall the main results here.\vspace{2mm}

{\it Direct basis:}\ Let $N$ be fixed and assume the convenient parametrization of the boundary parameters (\ref{param}). Define the family of $2^N$ possible  spins configurations 
$(\epsilon^{[i]}_1=\pm 1,\epsilon^{[i]}_2=\pm 1...,\epsilon^{[i]}_{N-1}=\pm 1,\epsilon^{[i]}_N=\pm 1)$
characterized by the quantum numbers $n,i$ such that\,\footnote{Given $n$, the ordering of states choosen here will be described below.}
\beqa
n=(N-\sum_{k=1}^{N}\epsilon^{[i]}_k)/2 \qquad \mbox{i.e.} \qquad 0 \leq n\leq N \qquad \mbox{and} \qquad i\in\{1,...,\left({{ N }\atop {n}}\right)\}\ \label{dom} .
\eeqa
In the canonical basis $\bigotimes_{j=1}^{N}|\pm\rangle_{j}$ on which the nonlocal operators (\ref{rep}) act, let us introduce the $2^N$ states 
\beqa
\psi^{(N)}_{n[i]}=\bigotimes_{l=1}^N \left( e^{\epsilon^{[i]}_l\alpha+\epsilon^{[i]}_l\sum_{k=1}^{l-1}\epsilon^{[i]}_k\phi/2+i\theta}|+\rangle_{l} + |-\rangle_{l} \right)\ \label{eigenvectN}
\eeqa
with (\ref{dom}). For generic boundary parameters and $q$, these states form a complete\,\footnote{For special relations among the boundary parameters, the representation may become indecomposable. We do not consider such possibility here.} basis of ${\cal V}$ on which the tridiagonal pair ${\cal W}_{0}^{(N)}$, ${\cal W}_{1}^{(N)}$ acts \cite{TDpair}. Using the explicit expressions (\ref{rep}) it is possible to exhibit the diagonal and block tridiagonal structure of the tridiagonal pair ${\cal W}_{0}^{(N)}$, ${\cal W}_{1}^{(N)}$ in this new basis: 
\beqa
{\cal W}^{(N)}_0\psi^{(N)}_{n[j]}&=&\lambda^{(N)}_n\psi^{(N)}_{n[j]} \qquad \mbox{with} \qquad \lambda^{(N)}_n=\cosh(\alpha+(N-2n)\phi/2)\ ,\nonumber\\
{\cal W}^{(N)}_1\psi^{(N)}_{n[j]}&=& \sum_{i=1}^{\binombN}\bcal^{(N,0)}_{n[ij]}\psi^{(N)}_{n+1[i]} + \sum_{i=1}^{\binomaN}\acal^{(N,0)}_{n[ij]}\psi^{(N)}_{n[i]} + \sum_{i=1}^{\binomcN}\ccal^{(N,0)}_{n[ij]}\psi^{(N)}_{n-1[i]} \ \label{recW1}
\eeqa
with $n=0,1,...,N$ and $j\in\{1,...,\big({{ N }\atop {n}}\big)\}$. Note that the explicit form of the coefficients $\acal^{(N,0)}_{n[ij]},\bcal^{(N,0)}_{n[ij]},\ccal^{(N,0)}_{n[ij]}$ - calculated in \cite{TDpair} - are reported in Appendix A.
According to (\ref{G1}), the nonlocal operators  ${\cal G}^{(N)}_1$ and ${\tilde{\cal G}}^{(N)}_1$ also possess a block tridiagonal structure in the eigenbasis (\ref{eigenvectN}) of ${\cal W}^{(N)}_0$. Plugging (\ref{recW1}) in (\ref{G1}), one finds
\beqa
{\cal G}^{(N)}_1\psi^{(N)}_{n[j]}&=& \ \ \sum_{i=1}^{\binombN}(q^{1/2}\lambda^{(N)}_{n}-q^{-1/2}\lambda^{(N)}_{n+1})\bcal^{(N,0)}_{n[ij]}\psi^{(N)}_{n+1[i]} + \sum_{i=1}^{\binomaN}(q^{1/2}-q^{-1/2})\lambda^{(N)}_{n}\acal^{(N,0)}_{n[ij]}\psi^{(N)}_{n[i]} \nonumber \\ 
&&+\ \sum_{i=1}^{\binomcN}(q^{1/2}\lambda^{(N)}_{n}-q^{-1/2}\lambda^{(N)}_{n-1})\ccal^{(N,0)}_{n[ij]}\psi^{(N)}_{n-1[i]}\ ,\ \nonumber\\
{\tilde{\cal G}}^{(N)}_1\psi^{(N)}_{n[j]}&=& \ \ \sum_{i=1}^{\binombN}(q^{1/2}\lambda^{(N)}_{n+1}-q^{-1/2}\lambda^{(N)}_{n})\bcal^{(N,0)}_{n[ij]}\psi^{(N)}_{n+1[i]} + \sum_{i=1}^{\binomaN}(q^{1/2}-q^{-1/2})\lambda^{(N)}_{n}\acal^{(N,0)}_{n[ij]}\psi^{(N)}_{n[i]} \nonumber \\ 
&&+\ \sum_{i=1}^{\binomcN}(q^{1/2}\lambda^{(N)}_{n-1}-q^{-1/2}\lambda^{(N)}_{n})\ccal^{(N,0)}_{n[ij]}\psi^{(N)}_{n-1[i]} \ .\nonumber
\eeqa

Proceeding similarly for higher values of $l=1,2,...,N-1$, it is an exercise to show that  {\it all} nonlocal operators defined by (\ref{rep}) {\it also} enjoy a block tridiagonal structure in the basis (\ref{eigenvectN}). To show it, it is useful to recall that the states (\ref{eigenvectN}) are ordered such that
\beqa
\psi^{(N)}_{n[j]}&=&\Big( e^{\alpha+(N-1-2n)\phi/2+i\theta}|+\rangle_{N} + |-\rangle_{N} \Big)\otimes \psi^{(N-1)}_{n[j]}\ \quad \quad\ \ \quad \quad\quad \mbox{for}\quad j\in \{1,...,\Big({{N-1}\atop {n}}\Big)\}\ ,\nonumber\\
\psi^{(N)}_{n[j]}&=&\Big(e^{-\alpha-(N+1-2n)\phi/2+i\theta}|+\rangle_{N} + |-\rangle_{N} \Big)\otimes \psi^{(N-1)}_{n-1[j-\binomNmun]}\ \quad \quad \ \mbox{for}\quad j\in \{\Big({{N-1}\atop {n}}\Big)+1,...,\Big({{N}\atop {n}}\Big)\}\ \nonumber
\eeqa
which are used to obtain recursive formulae for the coefficients of the block tridiagonal matrices. Using (\ref{rep}) together with the linear relations (\ref{c4}), after some straightforward calculations we finally get:
\beqa
{\cal W}^{(N)}_{-l}\psi^{(N)}_{n[j]}&=& \sum_{i=1}^{\binomaN}\dcal^{(N,l)}_{n[ij]}\psi^{(N)}_{n[i]}\ ,\nonumber\\ 
{\cal W}^{(N)}_{l+1}\psi^{(N)}_{n[j]}&=& \sum_{i=1}^{\binombN}\bcal^{(N,l)}_{n[ij]}\psi^{(N)}_{n+1[i]} + \sum_{i=1}^{\binomaN}\acal^{(N,l)}_{n[ij]}\psi^{(N)}_{n[i]} + \sum_{i=1}^{\binomcN}\ccal^{(N,l)}_{n[ij]}\psi^{(N)}_{n-1[i]}\ ,\nonumber\\
{\cal G}^{(N)}_{l+1}\psi^{(N)}_{n[j]}&=&\ \ \sum_{i=1}^{\binombN}(q^{1/2}\lambda^{(N)}_{n}-q^{-1/2}\lambda^{(N)}_{n+1})\bcal^{(N,l)}_{n[ij]}\psi^{(N)}_{n+1[i]} + \sum_{i=1}^{\binomaN}\hcal^{(N,l)}_{n[ij]}\psi^{(N)}_{n[i]} \nonumber \\ 
&&+\ \sum_{i=1}^{\binomcN}(q^{1/2}\lambda^{(N)}_{n}-q^{-1/2}\lambda^{(N)}_{n-1})\ccal^{(N,l)}_{n[ij]}\psi^{(N)}_{n-1[i]}\ ,\nonumber\\
{\tilde{\cal G}}^{(N)}_{l+1}\psi^{(N)}_{n[j]}&=&\ \ \sum_{i=1}^{\binombN}(q^{1/2}\lambda^{(N)}_{n+1}-q^{-1/2}\lambda^{(N)}_{n})\bcal^{(N,l)}_{n[ij]}\psi^{(N)}_{n+1[i]} + \sum_{i=1}^{\binomaN}\hcal^{(N,l)}_{n[ij]}\psi^{(N)}_{n[i]} \nonumber \\ 
&&+\ \sum_{i=1}^{\binomcN}(q^{1/2}\lambda^{(N)}_{n-1}-q^{-1/2}\lambda^{(N)}_{n})\ccal^{(N,l)}_{n[ij]}\psi^{(N)}_{n-1[i]}\label{recgen}
\eeqa
where all coefficients are reported in the Appendix A. Note that they have a rather simple form regarding to the recursion on the index $l$, a fact which may have been anticipated from the linear relations (\ref{c4}).\vspace{2mm}

{\it Dual basis:} The previous analysis can also be conducted considering the eigenbasis of ${\cal W}^{(N)}_{1}$ introduced in \cite{TDpair} instead of ${\cal W}^{(N)}_{0}$, leading to analogous results. Due to their close relationship with the expressions derived above, we only report the main results here. Recall that the $2^N$ states defined by
\beqa
{\varphi}^{(N)}_{s[k]}=\psi^{(N)}_{s[k]}|_{\alpha\rightarrow -\alpha^*,\ \phi\rightarrow - \phi,\ \theta\rightarrow \theta+i\pi}\label{phibasis} 
\eeqa
form an eigenbasis of ${\cal W}^{(N)}_{1}$ \cite{TDpair}. In this basis, the action of all nonlocal operators reads:
\beqa
{\cal W}^{(N)}_{-l}\varphi^{(N)}_{s[k]}&=&\sum_{m=1}^{\binomNspun}\bcalt^{(N,l)}_{s[mk]}\varphi^{(N)}_{s+1[m]} + \sum_{m=1}^{\binomNs}\acalt^{(N,l)}_{s[mk]}\varphi^{(N)}_{s[m]} + \sum_{m=1}^{\binomNsmun}\ccalt^{(N,l)}_{s[mk]}\varphi^{(N)}_{s-1[m]}\ ,\nonumber\\
{\cal W}^{(N)}_{l+1}\varphi^{(N)}_{s[k]}&=&\sum_{m=1}^{\binomNs}\dcalt^{(N,l)}_{s[mk]}\varphi^{(N)}_{s[m]}\ ,\nonumber\\ 
{\cal G}^{(N)}_{l+1}\varphi^{(N)}_{s[k]}&=&\ \ \sum_{m=1}^{\binomNspun}(q^{1/2}\lambda^{(N)}_{s+1}-q^{-1/2}\lambda^{(N)}_{s})\bcalt^{(N,l)}_{s[mk]}\varphi^{(N)}_{s+1[m]} + \sum_{m=1}^{\binomNs}\hcalt^{(N,l)}_{s[mk]}\varphi^{(N)}_{s[m]} \nonumber \\  &&+\ \sum_{m=1}^{\binomNsmun}(q^{1/2}\lambda^{(N)}_{s-1}-q^{-1/2}\lambda^{(N)}_{s})\ccalt^{(N,l)}_{s[mk]}\varphi^{(N)}_{s-1[m]}
\ ,\nonumber\\
{\tilde{\cal G}}^{(N)}_{l+1}\varphi^{(N)}_{s[k]}&=&\ \ \sum_{m=1}^{\binomNspun}(q^{1/2}\lambda^{(N)}_{s}-q^{-1/2}\lambda^{(N)}_{s+1})\bcalt^{(N,l)}_{s[mk]}\varphi^{(N)}_{s+1[m]} + \sum_{m=1}^{\binomNs}\hcalt^{(N,l)}_{s[mk]}\varphi^{(N)}_{s[m]} \nonumber \\  &&+\ \sum_{m=1}^{\binomNsmun}(q^{1/2}\lambda^{(N)}_{s}-q^{-1/2}\lambda^{(N)}_{s-1})\ccalt^{(N,l)}_{s[mk]}\varphi^{(N)}_{s-1[m]}
\label{recgendual}
\eeqa
with the identifications
\beqa
\xcalt^{(N,l)}_{s[mk]}&\equiv&\xcal^{(N,l)}_{s[mk]}|_{\alpha\leftrightarrow -\alpha^*, \ \phi\rightarrow -\phi} \qquad \mbox{for}\qquad \xcal\in\{\acal,\bcal,\ccal,\dcal\}\ ,\nonumber\\
\hcalt^{(N,l)}_{s[mk]}&\equiv&-\hcal^{(N,l)}_{s[mk]}|_{\alpha\leftrightarrow -\alpha^*, \ \phi\rightarrow -\phi}\ .\nonumber
\eeqa

{\it Functional basis:} Our aim is now to describe the functional basis which provides another useful realization of the nonlocal operators ${\cal W}^{(N)}_{-l} ,{\cal W}^{(N)}_{l+1}, {\cal G}^{(N)}_{l+1}, {\tilde{\cal G}}^{(N)}_{l+1}$ and will play an important role in the derivation of the transfer matrix spectrum.
Using the scalar product $(.,.)$ in the canonical basis $\bigotimes_{j=1}^{N}|\pm\rangle_{j}$, let us consider\footnote{The normalization is different from \cite{TDpair}.} the family of elements in the discrete argument $z_s=\exp(\alpha^*+(N-2s)\varphi/2)$:
\beqa
{\textsf F}^{(N)}_{n[jk]}(z_s)=\big({\tilde\varphi}^{(N)}_{s[k]}, \psi^{(N)}_{n[j]} \big) \qquad \mbox{with}\qquad n=0,1,...,N,\quad j\in \{1,...,\Big({{N}\atop {n}}\Big)\}\label{defFN}
\eeqa
where
\beqa
{\tilde\varphi}^{(N)}_{s[k]}=\bigotimes_{l=1}^N \left( e^{-\epsilon'^{[i]}_l\alpha+\epsilon'^{[i]}_l\sum_{m=1}^{l-1}\epsilon'^{[i]}_m\phi/2+i\theta}|+\rangle_{l} + |-\rangle_{l} \right)\ \quad \mbox{with}\quad s=(N-\sum_{m=1}^{N}\epsilon'^{[i]}_m)/2\label{dualeigen}
\eeqa
for any $s=0,1,...,N$ and $k\in \{\big(1,...,\big({{N}\atop {s}}\big)\}$. Given $k,s$ fixed, there are exactly $2^N$ elements of this form which are easily computed by taking the scalar product of (\ref{eigenvectN}) with (\ref{dualeigen}).\vspace{1mm} 

Having in mind that $({\cal W}^{(N)}_0)^\dagger\equiv {\cal W}^{(N)}_1$ as $\phi^\dagger=-\phi$ and $\alpha^\dagger=\alpha^*$, one writes the obvious identitities
\beqa
\big({\tilde\varphi}^{(N)}_{s[k]}, {\cal W}_{0}^{(N)}\psi^{(N)}_{n[i]}\big)= \big({\cal W}^{(N)}_1{\tilde\varphi}^{(N)}_{s[k]},\psi^{(N)}_{n[i]}\big) \qquad \mbox{and} \qquad \big({\cal W}^{(N)}_0{\tilde\varphi}^{(N)}_{s[k]},\psi^{(N)}_{n[i]}\big)=\big({\tilde\varphi}^{(N)}_{s[k]}, {\cal W}_{1}^{(N)}\psi^{(N)}_{n[i]}\big) \ .\label{identi}
\eeqa 
Acting with the nonlocal operators ${\cal W}^{(N)}_0,{\cal W}^{(N)}_1$ on the basis (\ref{eigenvectN}), (\ref{dualeigen}) one can show that the family of elements $\{{\textsf F}^{(N)}_{n[jk]}(z_s)\}$ satisfy a system of coupled three-term recurrence relations in $n$ and a system of coupled second-order discrete $q-$difference equations in the argument $z_s$ of the form (see \cite{TDpair} for details)
\beqa
\lambdat^{(N)}_s{\textsf F}^{(N)}_{n[jk]}(z_s)&=& \sum_{i=1}^{\binombN}\bcal^{(N,0)}_{n[ij]}{\textsf F}^{(N)}_{n+1[ik]}(z_s) + \sum_{i=1}^{\binomaN}\acal^{(N,0)}_{n[ij]}{\textsf F}^{(N)}_{n[ik]}(z_s) + \sum_{i=1}^{\binomcN}\ccal^{(N,0)}_{n[ij]}{\textsf F}^{(N)}_{n-1[ik]}(z_s)\ ,\label{recFN}\\
\lambda^{(N)}_n{\textsf F}^{(N)}_{n[jk]}(z_s)&=& \sum_{m=1}^{\binomNspun}\bcalt^{(N,0)}_{s[mk]}{\textsf F}^{(N)}_{n[jm]}(q^{-1}z_s) + \sum_{m=1}^{\binomNs}\acalt^{(N,0)}_{s[mk]}{\textsf F}^{(N)}_{n[jm]}(z_s) + \sum_{m=1}^{\binomNsmun}\ccalt^{(N,0)}_{s[mk]}{\textsf F}^{(N)}_{n[jm]}(qz_s)\ , \label{qdiff}
\eeqa
respectively. Here, note that we have introduced the ``dual'' spectrum $\lambdat^{(N)}_s=(z_s + z_s^{-1})/2$.\vspace{1mm} 

As explained in \cite{TDpair}, the equations (\ref{recFN}), (\ref{qdiff}) generalize the ones defining the Askey-Wilson $q-$orthogonal polynomials on a discret support. The main difference here is the presence of degeneracies in the spectrum of ${\cal W}^{(N)}_0$, ${\cal W}^{(N)}_1$: for $N>1$, the family $\{{\textsf F}^{(N)}_{n[jk]}(z_s)\}$ is associated with rational symmetric functions of the discrete argument $\lambdat^{(N)}_s$ (symmetric under the change $z_s\rightarrow z_s^{-1}$) with $s=0,1,...,N$. Associated with each equation in (\ref{identi}), from (\ref{recFN}), (\ref{qdiff}) one identifies the action of ${\cal W}^{(N)}_0$, ${\cal W}^{(N)}_1$ on the space of functions $\{{\textsf F}^{(N)}_{n[jk]}(z_s)\}$, as well as the action of ${\cal G}^{(N)}_1$, ${\tilde{\cal G}}^{(N)}_1$. In particular, the operator	associated with ${\cal W}^{(N)}_0$ can be seen as a generalization of the second-order $q-$difference Askey-Wilson operator \cite{TDpair}.\vspace{1mm} 

Generalizing the calculations to higher values of $l=1,2,..,N-1$, it is straightforward to show that any of the nonlocal operator ${\cal W}^{(N)}_{-l} ,{\cal W}^{(N)}_{l+1}, {\cal G}^{(N)}_{l+1}, {\tilde{\cal G}}^{(N)}_{l+1}$ acts on the functional basis (\ref{defFN}) as a second-order discrete $q-$difference operator in the argument $z_s$. We eventually get
\beqa
\big({\tilde\varphi}^{(N)}_{s[k]}, {\cal W}_{-l}^{(N)}\psi^{(N)}_{n[j]}\big)
&=& \sum_{m=1}^{\binomNspun}\bcalt^{(N,l)}_{s[mk]}{\textsf F}^{(N)}_{n[jm]}(q^{-1}z_s) + \sum_{m=1}^{\binomNs}\acalt^{(N,l)}_{s[mk]}{\textsf F}^{(N)}_{n[jm]}(z_s) + \sum_{m=1}^{\binomNsmun}\ccalt^{(N,l)}_{s[mk]}{\textsf F}^{(N)}_{n[jm]}(qz_s)\ ,\nonumber\\
\big({\tilde\varphi}^{(N)}_{s[k]}, {\cal W}_{l+1}^{(N)}\psi^{(N)}_{n[j]}\big)
&=& \sum_{m=1}^{\binomNs}\dcalt^{(N,l)}_{s[mk]}{\textsf F}^{(N)}_{n[jm]}(z_s)\ ,\nonumber\\
\big({\tilde\varphi}^{(N)}_{s[k]}, {\cal G}_{l+1}^{(N)}\psi^{(N)}_{n[j]}\big)
&=& \ \ \sum_{m=1}^{\binomNspun}(q^{1/2}\lambda^{(N)}_{s}-q^{-1/2}\lambda^{(N)}_{s+1})\bcalt^{(N,l)}_{s[mk]}{\textsf F}^{(N)}_{n[jm]}(q^{-1}z_s) + \sum_{m=1}^{\binomNs}\hcalt^{(N,l)}_{s[mk]}{\textsf F}^{(N)}_{n[jm]}(z_s)\nonumber\\
&& + \sum_{m=1}^{\binomNsmun}(q^{1/2}\lambda^{(N)}_{s}-q^{-1/2}\lambda^{(N)}_{s-1})\ccalt^{(N,l)}_{s[mk]}{\textsf F}^{(N)}_{n[jm]}(qz_s)\ ,\nonumber\\
\big({\tilde\varphi}^{(N)}_{s[k]}, {\tilde{\cal G}}_{l+1}^{(N)}\psi^{(N)}_{n[j]}\big)
&=& \ \ \sum_{m=1}^{\binomNspun}(q^{1/2}\lambda^{(N)}_{s+1}-q^{-1/2}\lambda^{(N)}_{s})\bcalt^{(N,l)}_{s[mk]}{\textsf F}^{(N)}_{n[jm]}(q^{-1}z_s) + \sum_{m=1}^{\binomNs}\hcalt^{(N,l)}_{s[mk]}{\textsf F}^{(N)}_{n[jm]}(z_s)\nonumber\\
&& + \sum_{m=1}^{\binomNsmun}(q^{1/2}\lambda^{(N)}_{s-1}-q^{-1/2}\lambda^{(N)}_{s})\ccalt^{(N,l)}_{s[mk]}{\textsf F}^{(N)}_{n[jm]}(qz_s)\ .\label{qdiffgen} 
\eeqa

Alternatively, it is worth interesting to notice that we also have
\beqa
\big({\tilde\psi}^{(N)}_{n[j]}, \varphi^{(N)}_{s[k]} \big)={\textsf F}^{(N)}_{n[jk]}(z_s) \qquad \mbox{for}\qquad {\tilde{\psi}}^{(N)}_{n[j]}=\varphi^{(N)}_{n[j]}|_{\alpha^*\rightarrow -\alpha^*}\ .\label{psitildebasis} 
\eeqa
As a consequence, using (\ref{recgen}) it is alternatively possible to derive a dual realization of the nonlocal operators acting on the functional basis ${\textsf F}^{(N)}_{n[jk]}(z_s)$. Although we will use it in the next Section, we leave it as an exercise for the reader.\vspace{1mm}

Now, having in hand all useful realizations (\ref{recgen}), (\ref{recgendual}), (\ref{qdiffgen}) - and its dual version just mentionned -  of the nonlocal operators ${\cal W}^{(N)}_{-l} ,{\cal W}^{(N)}_{l+1}, {\cal G}^{(N)}_{l+1}, {\tilde{\cal G}}^{(N)}_{l+1}$, we now turn to the block tridiagonal structure of the transfer matrix (\ref{tfin}) in each basis.

\subsection{Block tridiagonal structure of the transfer matrix}
From (\ref{Ifin}), the action of the nonlocal mutually commuting operators ${\cal I}^{(N)}_{2l+1}$ in above basis readily follows. For generic boundary conditions, none of the entries of the block tridiagonal matrices are vanishing and all states $\{\psi^{(N)}_{n[i]}\}$ are linearly independent. Then,
\beqa
{\cal I}^{(N)}_{2l+1}\psi^{(N)}_{n[j]}&=& \sum_{i=1}^{\binombN}\Bcal^{(N,l)}_{n[ij]}\psi^{(N)}_{n+1[i]} + \sum_{i=1}^{\binomaN}\Acal^{(N,l)}_{n[ij]}\psi^{(N)}_{n[i]} + \sum_{i=1}^{\binomcN}\Ccal^{(N,l)}_{n[ij]}\psi^{(N)}_{n-1[i]}\ ,\label{actionIn}\\
\big({\tilde\varphi}^{(N)}_{s[k]}, {\cal I}_{2l+1}^{(N)}\psi^{(N)}_{n[j]}\big)
&=& \sum_{m=1}^{\binomNspun}\Bcalt^{(N,l)}_{s[mk]}{\textsf F}^{(N)}_{n[jm]}(q^{-1}z_s) + \sum_{m=1}^{\binomNs}\Acalt^{(N,l)}_{s[mk]}{\textsf F}^{(N)}_{n[jm]}(z_s) + \sum_{m=1}^{\binomNsmun}\Ccalt^{(N,l)}_{s[mk]}{\textsf F}^{(N)}_{n[jm]}(qz_s)\ ,\label{actionIs}
\eeqa
where the coefficients read
\beqa
{\Bcal}^{(N,l)}_{n[ij]}&=&\left(\kappa^*+  \frac{\kappa_+}{k_+}\big(q^{1/2}\lambda^{(N)}_n-q^{-1/2}\lambda^{(N)}_{n+1}\big) + \frac{\kappa_-}{k_-}\big(q^{1/2}\lambda^{(N)}_{n+1}-q^{-1/2}\lambda^{(N)}_{n}\big)\right)\bcal^{(N,l)}_{n[ij]}\ ,\nonumber\\ 
{\Ccal}^{(N,l)}_{n[ij]}&=&\left( \kappa^*+  \frac{\kappa_+}{k_+}\big(q^{1/2}\lambda^{(N)}_n-q^{-1/2}\lambda^{(N)}_{n-1}\big) + \frac{\kappa_-}{k_-}\big(q^{1/2}\lambda^{(N)}_{n-1}-q^{-1/2}\lambda^{(N)}_{n}\big)    \right)\ccal^{(N,l)}_{n[ij]}\ ,\nonumber\\ 
{\Acal}^{(N,l)}_{n[ij]}&=&\kappa\dcal^{(N,l)}_{n[ij]} + \kappa^*\acal^{(N,l)}_{n[ij]} + \left(\frac{\kappa_+}{k_+}+\frac{\kappa_-}{k_-}\right)\hcal^{(N,l)}_{n[ij]}\ \label{Xn}
\eeqa 
and 
\beqa
\Xcalt^{(N,l)}_{s[mk]} = \Xcal^{(N,l)}_{s[mk]}|_{\kappa \leftrightarrow \kappa^*;\xcal\rightarrow \xcalt;\lambda\rightarrow \lambdat} \quad\mbox{for}\quad \xcal\in\{\acal,\bcal,\ccal,\dcal,\hcal\}\quad  \mbox{and} \quad \Xcal\in\{\Acal,\Bcal,\Ccal\}\ .\nonumber
\eeqa

Alternatively, from the action of all nonlocal operators on the dual basis (\ref{phibasis}) one immediately gets
\beqa
{\cal I}^{(N)}_{2l+1}\varphi^{(N)}_{s[k]}&=& \sum_{m=1}^{\binomNspun}{\Bcalt}^{'(N,l)}_{s[mk]}\varphi^{(N)}_{s+1[m]} + \sum_{m=1}^{\binomNs}\Acalt^{'(N,l)}_{s[mk]}\varphi^{(N)}_{s[m]} + \sum_{m=1}^{\binomNsmun}\Ccalt^{'(N,l)}_{s[mk]}\varphi^{(N)}_{s-1[m]}\ ,\label{actionIsdual}\\
\big({\tilde\psi}^{(N)}_{n[j]}, {\cal I}_{2l+1}^{(N)}\varphi^{(N)}_{s[k]}\big)
&=& \sum_{i=1}^{\binombN}\Bcal^{'(N,l)}_{n[ij]}{\textsf F}^{(N)}_{n+1[im]}(z_s) + \sum_{i=1}^{\binomaN}\Acal^{'(N,l)}_{n[ij]}{\textsf F}^{(N)}_{n[im]}(z_s) + \sum_{i=1}^{\binomcN}\Ccal^{'(N,l)}_{n[ij]}{\textsf F}^{(N)}_{n-1[im]}(z_s)\ ,\label{actionIndual}
\eeqa
where the ``dual'' coefficients read
\beqa
{\Bcal}^{'(N,l)}_{s[mk]}&=&\left(\kappa+  \frac{\kappa_+}{k_+}\big(q^{1/2}\lambdat^{(N)}_{s+1}-q^{-1/2}\lambdat^{(N)}_{s}\big) + \frac{\kappa_-}{k_-}\big(q^{1/2}\lambdat^{(N)}_{s}-q^{-1/2}\lambdat^{(N)}_{s+1}\big)\right)\bcalt^{(N,l)}_{s[mk]}\ ,\nonumber\\ 
{\Ccal}^{'(N,l)}_{s[mk]}&=&\left( \kappa+  \frac{\kappa_+}{k_+}\big(q^{1/2}\lambdat^{(N)}_{s-1}-q^{-1/2}\lambdat^{(N)}_{s}\big) + \frac{\kappa_-}{k_-}\big(q^{1/2}\lambdat^{(N)}_{s}-q^{-1/2}\lambdat^{(N)}_{s-1}\big)    \right)\ccalt^{(N,l)}_{s[mk]}\ ,\nonumber\\ 
{\Acal}^{'(N,l)}_{s[mk]}&=&\kappa\acalt^{(N,l)}_{s[mk]} + \kappa^*\dcalt^{(N,l)}_{s[mk]} + \left(\frac{\kappa_+}{k_+}+\frac{\kappa_-}{k_-}\right)\hcalt^{(N,l)}_{s[mk]}\ \label{Xprimes}
\eeqa 
and 
\beqa
\Xcalt^{'(N,l)}_{n[ij]} = \Xcal^{'(N,l)}_{n[ij]}|_{\kappa^* \leftrightarrow \kappa;\xcalt\rightarrow \xcal;\lambdat\rightarrow \lambda} \quad\mbox{for}\quad \xcalt\in\{\acalt,\bcalt,\ccalt,\dcalt,\hcalt\}\quad  \mbox{and} \quad \Xcalt\in\{\Acalt,\Bcalt,\Ccalt\}\ .\nonumber
\eeqa

This demonstrates the remarkable block tridiagonal structure of the transfer matrix (\ref{tfin}) in each basis. In particular, from the quantity $\big({\tilde\varphi}^{(N)}_{s[k]},t_{XXZ}(u)\psi^{(N)}_{n[i]}\big)$ (or alternatively $\big({\tilde\psi}^{(N)}_{n[i]},t_{XXZ}(u){\varphi}^{(N)}_{s[k]}\big)$) one extracts a generating function for a family of mutually commuting second-order discrete $q-$difference operators generalizing the Askey-Wilson one.

\section{Eigenstates and spectrum of the XXZ open spin chain}
\subsection{Generic integrable boundary conditions}
For generic values of the boundary parameters and deformation parameter $q$, none of the coefficients (\ref{Xn}) are vanishing. In other words, there is no nontrivial subspace of $\cal V$ which is left invariant under the action of the nonlocal commuting operators ${\cal I}^{(N)}_{2l+1}$. Then, any common eigenstate of all nonlocal commuting operators ${\cal I}^{(N)}_{2l+1}$ can be written {\it either} in the eigenbasis of ${\cal W}^{(N)}_{0}$ - i.e. the direct basis (\ref{eigenvectN}) - {\it or} in the eigenbasis of ${\cal W}^{(N)}_{1}$ - i.e. the dual basis (\ref{phibasis}). Choosing for instance the most general linear combinations of states (\ref{eigenvectN}) any eigenstate of ${\cal I}^{(N)}_{2l+1}$ can be written
\beqa
\Psi^{(N)}_{\Lambda^{(N)}_1}= \sum_{n=0}^{N}\sum_{j=1}^{\big({{N}\atop {n}}\big)} f^{(N)}_{n[j]}\big(\Lambda^{(N)}_1\big)\psi^{(N)}_{n[j]}\ ,  \label{vectN}
\eeqa
where $\Lambda^{(N)}_1$ is an argument used to classify all posssible eigenstates of the transfer matrix (\ref{tfin}).
In view of (\ref{commut}) and the block tridiagonal structure of ${\cal W}^{(N)}_{1}$ (resp. ${\cal W}^{(N)}_{0}$ in the dual basis) with respect to the eigenbasis of ${\cal W}^{(N)}_{0}$ (resp. ${\cal W}^{(N)}_{1}$) for generic boundary parameters, it is actually sufficient\,\footnote{For generic values of the parameters, the spectrum of ${\cal I}^{(N)}_{1}$ does not have degeneracies.} to consider the action of  ${\cal I}^{(N)}_{1}$ on (\ref{vectN}) to determine all functions $f^{(N)}_{n[j]}\big(\Lambda^{(N)}_1\big)$ as well as the discrete support which $\Lambda^{(N)}_1$ belongs to. Using (\ref{actionIn}) for $l=0$, the $2^N$ functions are found to satisfy the system of coupled three-term recurrence relations on $n$ (see also \cite{qBethe}):
\beqa
\sum_{m=1}^{\big({{N}\atop {n-1}}\big)}{\cal B}^{(N,0)}_{n-1[jm]}f^{(N)}_{n-1[m]}\big(\Lambda^{(N)}_1\big) +\sum_{m=1}^{\big({{N}\atop {n+1}}\big)}{\cal C}^{(N,0)}_{n+1[jm]}f^{(N)}_{n+1[m]}\big(\Lambda^{(N)}_1\big)
+ \sum_{m=1}^{\big({{N}\atop {n}}\big)}\big({\cal A}^{(N,0)}_{n[jm]} - \Lambda^{(N)}_1\delta_{jm}\big)f^{(N)}_{n[m]}\big(\Lambda^{(N)}_1\big) =0\   \label{recf}
\eeqa
for $n=0,1,...,N$ and $j=1,...,\big({{N}\atop {n}}\big)$ with (\ref{Xn}), ${\cal B}^{(N,0)}_{-1[jm]}={\cal C}^{(N,0)}_{N+1[jm]}\equiv 0$ \ and we have defined
\beqa
{\cal I}^{(N)}_{1}\Psi^{(N)}_{\Lambda^{(N)}_1}=\Lambda^{(N)}_1\Psi^{(N)}_{\Lambda^{(N)}_1} \ .\nonumber
\eeqa
Appart from the occurence of degeneracies $j,m$ which introduce an additional complexity, such kind of equations often arises in the theory of $q-$orthogonal polynomials, and can be easily solved along the same lines. Normalizing $f_{0[1]}\big(\Lambda^{(N)}_1\big)\equiv 1$, $2^N-1$ equations of (\ref{recf}) determine uniquely the hierarchy of {\it rational} functions in the (yet unrestricted) argument  $\Lambda^{(N)}_1$. In addition, it is easy to see that the remaining equation coincides with ${\cal P}^{(N)}\big(\Lambda^{(N)}_1\big)=0$ \ where
\beqa
{\cal P}^{(N)}\big(\Lambda^{(N)}_1\big)=\det[{\cal I}^{(N)}_{1}-\Lambda^{(N)}_1I\!\!I^{(N)}]\label{P}
\eeqa
is the characteristic polynomial of ${\cal I}^{(N)}_{1}$ written in terms of the coefficients (\ref{Xn}). It determines the set of ``allowed'' eigenvalues\,\footnote{The spectral problem for the trivial special case $N=1$ of the XXZ open spin chain (\ref{H}) has some analogies with the one for the Azbel-Hofstadter model. From (\ref{recf}) one immediatly finds $f^{(1)}_{0[1]}(\Lambda^{(1)}_1)\equiv 1\ ,\quad f^{(1)}_{1[1]}(\Lambda^{(1)}_1)=-({\Acal}^{(1,0)}_{0[11]}-\Lambda^{(1)}_1)/{\Ccal}^{(1,0)}_{1[11]}\ $\ 
and the characteristic polynomial  ${\cal P}^{(1)}(\Lambda^{(1)}_1)= ({\Acal}^{(1,0)}_{0[11]}-\Lambda^{(1)}_1)({\Acal}^{(1,0)}_{1[11]}-\Lambda^{(1)}_1) - {{\Bcal}^{(1,0)}_{0[11]}}{{\Ccal}^{(1,0)}_{1[11]}}$ which has exactly two distinct roots.}:
\beqa
{\cal P}^{(N)}\big(\Lambda^{(N)}_1\big)=0 \qquad \Longleftrightarrow \qquad \Lambda^{(N)}_1\in\{\Lambda^{(N)}_{1,1},...,\Lambda^{(N)}_{1,2^N}\}\ .  \label{eigenv}
\eeqa

Consequently, the complete set of $2^N$ eigenstates of the XXZ open spin chain (\ref{H}) with generic boundary parameters and $q$ is given by
(\ref{vectN}) with (\ref{recf}), (\ref{eigenv}). Note that the completeness of the basis (\ref{vectN}) is a natural consequence of the dimension and irreducibility property of ${\cal V}$ for generic boundary parameters. In our approach, the counting of eigenstates is encoded in the degree $d$ of the characteristic polynomial (\ref{P}). For generic boundary conditions, the irreducibility of the representations (\ref{eigenvectN}), (\ref{vectN}) implies $d=2^N$.\vspace{1mm}

Let us now turn to the calculation of the eigenvalues of all nonlocal commuting operators ${\cal I}^{(N)}_{2l+1}$. The general spectral problem reads:
\beqa
{\cal I}^{(N)}_{2l+1}\Psi^{(N)}_{\Lambda^{(N)}_1}=\Lambda^{(N)}_{2l+1}\Psi^{(N)}_{\Lambda^{(N)}_1} \quad \ \mbox{for}\qquad l=0,1,...,N-1\ . \nonumber
\eeqa
Taking for instance\,\footnote{Alternatively, on may simply consider each of the recurrence relations associated with $l\geq1$ generalizing (\ref{recf}) for a given value of $n$. In this case, the eigenvalues are written in terms of the coefficients (\ref{Xn}) and ratios of $f^{(N)}_{n[j]}\big(\Lambda^{(N)}_1\big)$.} the scalar product of each side of the equation with (\ref{dualeigen}) and using (\ref{actionIs}), we end up with a set of second-order discrete $q-$difference equations on $s$ which - independently of each other - can be used to determine the eigenvalues. These constraints, satisfied for any $s=0,1,...,N$ and $k=1,...,\big({{N}\atop {s}}\big)$, read
\beqa
\Lambda^{(N)}_{2l+1}=   \sum_{m=1}^{\binomNspun}\Bcalt^{(N,l)}_{s[mk]}\frac{{\Psi}^{(N)}_{[m]}(s+1)}{{\Psi}^{(N)}_{[k]}(s)} + \sum_{m=1}^{\binomNs}\Acalt^{(N,l)}_{s[mk]}\frac{{\Psi}^{(N)}_{[m]}(s)}{{\Psi}^{(N)}_{[k]}(s)} + \sum_{m=1}^{\binomNsmun}\Ccalt^{(N,l)}_{s[mk]}\frac{{\Psi}^{(N)}_{[m]}(s-1)}{{\Psi}^{(N)}_{[k]}(s)}\ \label{spectrumI}
\eeqa
where the $2^N-$different basis of functions
\beqa
{\Psi}^{(N)}_{[k]}(s)= \sum_{n=0}^N\sum_{j=1}^{\big({{N}\atop {n}}\big)} f^{(N)}_{n[j]}\big(\Lambda^{(N)}_1\big){\textsf F}^{(N)}_{n[jk]}(z_s)\label{eigenf1}
\eeqa
with (\ref{defFN}) have been introduced. For generic parameters, it turns out that the eigenvalues $\Lambda_{XXZ}(u)$ of the transfer matrix (\ref{tfin}) are {\it rational} functions of the spectral parameter $u$ and $\Lambda^{(N)}_{1}$ defined on the discrete support (\ref{eigenv}). They read
\beqa
\Lambda_{XXZ}(u)=\sum_{l=0}^{N-1}{\cal F}_{2l+1}(u)\ \Lambda^{(N)}_{2l+1}  + {\cal F}_0(u)\ .\label{spectrum}
\eeqa
The spectrum $E$ of the Hamiltonian (\ref{H}) immediatly follows:
\beqa
E&=& \frac{(q^{1/2}-q^{-1/2})(q^{1/2}+q^{-1/2})^{-1}}{2(\kappa+\kappa^*)(\epsilon_+^{(0)}+\epsilon_-^{(0)})}\left(\sum_{l=0}^{N-1}\frac{d{\cal F}_{2l+1}(u)}{du}|_{u=1}\ \Lambda^{(N)}_{2l+1} + \frac{d{\cal F}_0(u)}{du}|_{u=1}\right)\ \nonumber \\
&&- \left(N\Delta + \frac{(q^{1/2}-q^{-1/2})^2}{2(q^{1/2}+q^{-1/2})}\right)\ .  \label{E}
\eeqa

\subsection{Special integrable boundary conditions and Bethe-type solutions}
For special relations between left and right boundary parameters, invariant nontrivial subspaces denoted ${\cal W}_\pm\subset {\cal V}$ with dim$({\cal W}_\pm)<$ dim$({\cal V})$ and ${\overline{\cal W}}_\pm\subset {\cal V}$ dim$(\overline{{\cal W}}_\pm)<$ with dim$({\cal V})$ such that 
\beqa
\mbox{dim}({\cal W}_+) + \mbox{dim}({{\cal W}}_-)&=& \mbox{dim}({\cal V})=2^N\ ,\nonumber\\
\mbox{dim}({\overline{\cal W}}_+) + \mbox{dim}(\overline{{\cal W}}_-)&=& \mbox{dim}({\cal V})=2^N\label{reldim}
\label{reldim}
\eeqa
can be constructed. Indeed, due to the block tridiagonal structure (\ref{actionIn}) or (\ref{actionIsdual}) of all nonlocal operators with respect to the eigenbasis of ${\cal W}_0^{(N)}$ and ${\cal W}_1^{(N)}$ for some vanishing entries in (\ref{Xn}) or (\ref{Xprimes}) it is possible to construct eigenstates of ${\cal I}^{(N)}_{2l+1}$ as truncated - compared with (\ref{vectN}) - linear combinations of states (\ref{eigenvectN}) or (\ref{phibasis}). Then, as we now show given a family of boundary conditions the complete set of eigenstates of the model (\ref{H}) split in two sectors:\vspace{1mm}

{\it Boundary conditions associated with ${\cal W}_\pm$:} Given an integer $0\leq p< N$ fixed, consider the two families of states written in the direct basis (\ref{eigenvectN}) and dual basis (\ref{phibasis}), respectively, as:
\beqa
\Psi^{(N,+)}_{\Lambda^{(N)}_1}= \sum_{n=0}^{p}\sum_{j=1}^{\big({{N}\atop {n}}\big)} f^{(N,+)}_{n[j]}\big(\Lambda^{(N)}_1\big)\psi^{(N)}_{n[j]}\ \qquad \mbox{and}\ \qquad {\Psi}^{(N,-)}_{\Lambda^{(N)}_1}= 
\sum_{s=0}^{N-p-1}\sum_{k=1}^{\big({{N}\atop {s}}\big)} f^{(N,-)}_{s[k]}\big(\Lambda^{(N)}_1\big)
\varphi^{(N)}_{s[k]}\ .\label{vecttrunc1}
\eeqa
According to (\ref{actionIn}) and (\ref{actionIsdual}), the states (\ref{vecttrunc1}) can span the two invariant subspaces denoted ${\cal W}_\pm$ provided the conditions 
\beqa
{\cal B}^{(N,l)}_{p[ij]}\equiv 0 \qquad \mbox{and} \qquad {\cal B}^{'(N,l)}_{N-p-1[mk]}\equiv 0\  \qquad \mbox{for all} \qquad l,i,j,m,k \nonumber
\eeqa
are satisfied, respectively. Written in terms of the boundary parameters and using the parametrization (\ref{param}) in (\ref{Xn}), (\ref{Xprimes}), it is easy to see that both conditions reduce to
\beqa
\ \alpha\pm{\tilde\alpha} &=& - i({\tilde \theta}-\theta) - (N-2p-1)\phi/2\qquad mod(2i\pi)\ . \label{r1}
\eeqa
In the first sector - associated with ${\cal W}_+$ -, the family of rational functions $f^{(N,+)}_{n[j]}\big(\Lambda^{(N)}_1\big)$ must satisfy the truncated (compared to (\ref{recf})) system of three-term recurrence relations
\beqa
\sum_{m=1}^{\big({{N}\atop {n-1}}\big)}{\cal B}^{(N,0)}_{n-1[jm]}f^{(N,+)}_{n-1[m]}\big(\Lambda^{(N)}_1\big) +\sum_{m=1}^{\big({{N}\atop {n+1}}\big)}{\cal C}^{(N,0)}_{n+1[jm]}f^{(N,+)}_{n+1[m]}\big(\Lambda^{(N)}_1\big)
+ \sum_{m=1}^{\big({{N}\atop {n}}\big)}\big({\cal A}^{(N,0)}_{n[jm]} - \Lambda^{(N)}_1\delta_{jm}\big)f^{(N,+)}_{n[m]}\big(\Lambda^{(N)}_1\big) =0\   \label{recfM}
\eeqa
with \ $n=0,1,...,p$. On the other hand, in the second sector - associated with ${\cal W}_-$ - one has 
\beqa
\sum_{m=1}^{\big({{N}\atop {s-1}}\big)}{\cal B}^{'(N,0)}_{s-1[km]}f^{(N,-)}_{s-1[m]}\big(\Lambda^{(N)}_1\big) +\sum_{m=1}^{\big({{N}\atop {s+1}}\big)}{\cal C}^{'(N,0)}_{s+1[km]}{ f}^{(N,-)}_{s+1[m]}\big(\Lambda^{(N)}_1\big)
+ \sum_{m=1}^{\big({{N}\atop {s}}\big)}\big({\cal A}^{'(N,0)}_{s[km]} - \Lambda^{(N)}_1\delta_{km}\big){ f}^{(N,-)}_{s[m]}\big(\Lambda^{(N)}_1\big) =0\   \label{recfMdual}
\eeqa
with \ $s=0,1,...,N-p-1$. Similarly to the case of generic boundary parameters, dim$({\cal W}_\pm)-1$  equations of (\ref{recfM}) or (\ref{recfMdual}) determine the functions $f^{(N,\pm)}_{n[m]}\big(\Lambda^{(N)}_1\big)$, whereas each of the two remaining equations coincides - up to an overall factor - with two conditions on some truncated characteristic polynomials
\beqa
{\cal P}^{(N,+)}\big(\Lambda^{(N)}_1\big)=0 &&\qquad \Longleftrightarrow \qquad \Lambda^{(N)}_1\in\{\Lambda^{(N)}_{1,1},...,\Lambda^{(N)}_{1,\mbox{dim}({\cal W}_+)}\}\ ,\label{eigenvPp}\\
{\cal P}^{(N,-)}\big(\Lambda^{(N)}_1\big)=0 &&\qquad \Longleftrightarrow \qquad \Lambda^{(N)}_1\in\{\Lambda^{(N)}_{1,2^N-\mbox{dim}({\cal W}_-)+1},...,\Lambda^{(N)}_{1,2^N}\}\label{eigenvPm}
\eeqa
where 
\beqa
{\cal P}^{(N,\pm)}\big(\Lambda^{(N)}_1\big)= \det[{\cal I}^{(N,\pm)}_{1\ trunc}-\Lambda^{(N)}_1I\!\!I^{(\mbox{dim}({\cal W}_\pm))}] \ \label{eigenv1}
\eeqa
and the two truncated block tridiagonal matrices ${\cal I}^{(N,\pm)}_{1\ trunc}$ which entries appear in (\ref{recfM}) and (\ref{recfMdual}), respectively, have been introduced.\vspace{1mm}

Each linearly independent set of eigenstates in (\ref{vecttrunc1}) form an eigenbasis for ${\cal I}^{(N,\pm)}_{1\ trunc}$ of ${\cal W}_\pm$ which dimensions\,\footnote{Similar results have been independently obtained from the representation theory of the two-boundary Temberley-Lieb algebra \cite{Ritt}.}, respectively, follow from the degeneracies of each eigenspace of ${\cal W}_0^{(N)}$ and ${\cal W}_1^{(N)}$ \cite{TDpair}. One has 
\beqa
\mbox{dim}({\cal W}_+)=\sum_{n=0}^{p}\biggl({{N}\atop {n}}\biggr)\qquad \qquad \mbox{and} \qquad \qquad \mbox{dim}({\cal W}_-)=\sum_{s=0}^{N-p-1}\biggl({{N}\atop {s}}\biggr)  \ \label{Wdim} 
\eeqa
which clearly satisfy (\ref{reldim}). Then, given the family of boundary conditions (\ref{r1}) with $0\leq p<N$ the complete set of eigenstates $(\pm)$ of the transfer matrix (\ref{tfin}) is given by (\ref{vecttrunc1}) with (\ref{eigenvPp}) for $(+)$ and (\ref{eigenvPm}) for $(-)$.\vspace{2mm}

{\it Boundary conditions associated with ${\overline{\cal W}}_\pm$:} Given an integer $0\leq p< N$ fixed, let us consider instead the two families of states  defined by:
\beqa
\Psi^{(N,+)}_{\Lambda^{(N)}_1}= \sum_{n=p+1}^{N}\sum_{j=1}^{\big({{N}\atop {n}}\big)} f^{(N,+)}_{n[j]}\big(\Lambda^{(N)}_1\big)\psi^{(N)}_{n[j]}\ \qquad \mbox{and}\ \qquad {\Psi}^{(N,-)}_{\Lambda^{(N)}_1}= 
\sum_{s=N-p}^{N}\sum_{k=1}^{\big({{N}\atop {s}}\big)} f^{(N,-)}_{s[k]}\big(\Lambda^{(N)}_1\big)
\varphi^{(N)}_{s[k]}\ .\label{vecttrunc2}
\eeqa
Following the same procedure as above, it is easy to see that these states can span the invariant subspaces denoted ${\overline{\cal W}}_\pm$ provided 
\beqa
{\cal C}^{(N,l)}_{p+1[ij]}\equiv 0 \qquad \mbox{and} \qquad {\cal C}^{'(N,l)}_{N-p[mk]}\equiv 0\  \qquad \mbox{for all} \qquad l,i,j,m,k \ .\nonumber
\eeqa
In terms of the boundary parameters, these constraints become:
\beqa
 \ \alpha\pm{\tilde\alpha} &=& \ \ i({\tilde \theta}-\theta) - (N-2p-1)\phi/2 \qquad  mod(2i\pi)\ .\label{r2}
\eeqa
The corresponding families of rational functions entering in (\ref{vecttrunc2}) satisfy the system of truncated three-term recurrence relations (\ref{recfM}) with $n=p+1,...,N$ and (\ref{recfMdual}) with $s=N-p,...,N$, respectively. From these latter relations, the two characteristic polynomials 
\beqa
{\overline{\cal P}}^{(N,\pm)}\big(\Lambda^{(N)}_1\big)= \det[{\overline{\cal I}}^{(N,\pm)}_{1\ trunc}-\Lambda^{(N)}_1I\!\!I^{(\mbox{dim}({\overline{\cal W}}_\pm))}] \ \label{eigenv1}
\eeqa
are easily exhibited. Here, by analogy with the previous case we have introduced the two truncated block tridiagonal matrices ${\overline{\cal I}}^{(N,\pm)}_{1\ trunc}$. For each family of states $(\pm)$ in (\ref{vecttrunc2}), it turns out that the eigenvalues $\Lambda^{(N)}_1$ belong to a discret support defined by:
\beqa
{\overline{\cal P}}^{(N,+)}\big(\Lambda^{(N)}_1\big)=0 &&\qquad \Longleftrightarrow \qquad \Lambda^{(N)}_1\in\{\Lambda^{(N)}_{1,1},...,\Lambda^{(N)}_{1,\mbox{dim}({\overline{\cal W}_+)}}\}\ ,\label{eigenvPp2}\\
{\overline{\cal P}}^{(N,-)}\big(\Lambda^{(N)}_1\big)=0 &&\qquad \Longleftrightarrow \qquad \Lambda^{(N)}_1\in\{\Lambda^{(N)}_{1,2^N-\mbox{dim}({\overline{\cal W}}_-)+1},...,\Lambda^{(N)}_{1,2^N}\} \ .\label{eigenvPm2}
\eeqa
In view of (\ref{vecttrunc2}), note that one has (\ref{reldim}) with
\beqa
\mbox{dim}({\overline{\cal W}}_+)=\sum_{n=p+1}^{N}\biggl({{N}\atop {n}}\biggr)\qquad \qquad \mbox{and} \qquad \qquad \mbox{dim}({\overline{\cal W}}_-)=\sum_{s=N-p}^{N}\biggl({{N}\atop {s}}\biggr)  \ .\label{Wbardim} 
\eeqa
Then, given the family of boundary conditions (\ref{r2}) with $0\leq p<N$ the complete set of eigenstates $(\pm)$ of the transfer matrix (\ref{tfin}) is given by (\ref{vecttrunc2}) with (\ref{eigenvPp2}) for $(+)$ and (\ref{eigenvPm2}) for $(-)$.\vspace{2mm}

The complete set of eigenvalues of the transfer matrix (\ref{tfin}) can now be calculated. For any choice of boundary conditions (\ref{r1}) or (\ref{r2}), as a consequence of the eigenspaces decomposition the eigenvalues of all nonlocal operators ${\cal I}^{(N)}_{2l+1}$ split in two sets. Introducing the truncated functions
\beqa
{\Psi}^{(N,+)}_{[k]}(s)=  \big({\tilde\varphi}^{(N)}_{s[k]},  \Psi^{(N,+)}_{\Lambda^{(N)}_1}\big)  \qquad \mbox{and} \qquad  {\Psi}^{(N,-)}_{[j]}(n)=  \big({\tilde\psi}^{(N)}_{n[j]},  \Psi^{(N,-)}_{\Lambda^{(N)}_1}\big) \ ,  \label{eigenf2}
\eeqa
by analogy with the case of generic boundary conditions using (\ref{actionIs}) and (\ref{actionIndual}) we end up with two independent sets of second-order discrete $q-$difference equations on $s$ and $n$ which determine the eigenvalues for any $s,n,k,j$. We obtain 
\beqa
\Lambda^{(N,+)}_{2l+1}&=&   \sum_{m=1}^{\binomNspun}\Bcalt^{(N,l)}_{s[mk]}\frac{{\Psi}^{(N,+)}_{[m]}(s+1)}{{\Psi}^{(N,+)}_{[k]}(s)} + \sum_{m=1}^{\binomNs}\Acalt^{(N,l)}_{s[mk]}\frac{{\Psi}^{(N,+)}_{[m]}(s)}{{\Psi}^{(N,+)}_{[k]}(s)} + \sum_{m=1}^{\binomNsmun}\Ccalt^{(N,l)}_{s[mk]}\frac{{\Psi}^{(N,+)}_{[m]}(s-1)}{{\Psi}^{(N,+)}_{[k]}(s)}\ \ , \label{spectrumIp}\\
\Lambda^{(N,-)}_{2l+1}&=&   \sum_{m=1}^{\binombN}\Bcalt^{'(N,l)}_{n[mj]}\frac{{\Psi}^{(N,-)}_{[m]}(n+1)}{{\Psi}^{(N,-)}_{[j]}(n)} + \sum_{m=1}^{\binomaN}\Acalt^{'(N,l)}_{n[mj]}\frac{{\Psi}^{(N,-)}_{[m]}(n)}{{\Psi}^{(N,-)}_{[j]}(n)} + \sum_{m=1}^{\binomcN}\Ccalt^{'(N,l)}_{n[mj]}\frac{{\Psi}^{(N,-)}_{[m]}(n-1)}{{\Psi}^{(N,-)}_{[j]}(n)}\ .
\label{spectrumIm}
\eeqa
All together, using (\ref{spectrum}) they provide the complete set of eigenvalues of the transfer matrix as well as the energy levels (\ref{E}) of the model (\ref{H}) for the family of boundary conditions (\ref{r1}), (\ref{r2}).\vspace{2mm} 

It is worth interesting to notice that for the boundary parameter spaces (\ref{r1}), (\ref{r2}), the transfer matrix (\ref{tfin}) is known to admit Bethe-type solutions \cite{Nepo,Cao,WL}. Indeed, acting with the gauge-transformed creation/annihilation operators $B(u),C(u)$ on two different pseudo-reference states $|\Omega^{\pm}\rangle$ (see \cite{Cao,WL} for details) the complete set of Bethe eigenstates are formally written
\beqa
\Psi(u_1,...,u_p)\equiv B(u_1)...B(u_p)|\Omega^{+}\rangle \qquad \mbox{and} \qquad  \Psi(u_1,...,u_p)\equiv C(u_1)...C(u_p)| \Omega^{-}\rangle \label{Bethe}
\eeqa
where $\{u_k\}$ denote the Bethe roots. From our analysis, given the family of boundary conditions (\ref{r1}) or (\ref{r2}) it follows that the eigenstates (\ref{Bethe}) admit an alternative description of the form (\ref{vecttrunc1}) or (\ref{vecttrunc2}), respectively, as rational functions of the discrete variable $\Lambda_1^{(N)}$. In particular, setting $p=0$ in (\ref{r1}), (\ref{r2})  one identifies the two reference states 
\beqa
|\Omega^{+}\rangle \rightarrow \psi_{0[1]}^{(N)} \qquad \mbox{and} \qquad |\Omega^{-}\rangle \rightarrow \varphi_{N[1]}^{(N)} 
\eeqa
in agreement with \cite{Cao,WL}. For generic values of $p$ and boundary conditions of the form (\ref{r1}), (\ref{r2}) the correspondence between (\ref{Bethe}) and (\ref{vecttrunc1}) or (\ref{vecttrunc2}) allows to derive exact relations between Bethe roots and roots of the characteristic polynomials ${\cal P}^{(N,\pm)}(\Lambda_1^{(N)}), {\overline{\cal P}}^{(N,\pm)}(\Lambda_1^{(N)})$. These relations will be considered elsewhere.

\section{Concluding remarks}
The approach initiated in \cite{qDG} and based on the representation theory of the $q-$Onsager algebra opens the possibility of studying massive continuum or lattice quantum integrable models from a point of view which possesses some advantages compared to standard - like algebraic or functional Bethe ansatz - approaches: analytic solution for generic boundary and anisotropy parameters; no ansatz required; rigorous proof of completeness.
Applied here to the XXZ open spin chain it has shown its efficiency for generic integrable boundary conditions in (\ref{H}) as well as for the special conditions (\ref{r1}), (\ref{r2}) for which the complete spectrum has been derived. No modifications of the formalism between the two regimes of boundary parameters have been necessary, the only changes being the structure of the eigenstates and related characteristic polynomial.

Another nice feature of the present approach relies on the fact that the eigenvalues of the transfer matrix for any integrable boundary conditions are expressed in terms of the roots of a characteristic polynomial of degree $d=2^N$ ($d\leq 2^N$ for (\ref{r1}), (\ref{r2}) which gives all subsets of the eigenvalues), similarly to the Ising \cite{Ons} and superintegrable chiral Potts models \cite{Potts,Davies}. For the regime of parameters for which the Bethe ansatz applies, one is reduced to solve a single {\it algebraic} equation instead of {\it highly transcendental} equations (Bethe equations) in the Bethe ansatz approach. Numerically, a detailed comparison between both approaches is an interesting problem, which however goes beyond the scope of this paper.  
In particular, in the {\it algebraic} sectors (\ref{r1}), (\ref{r2}) associated with Bethe-type solutions, it is worth noticing that such a correspondence between these roots and solutions of Bethe equations should not be surprising: it is a well-known fact in the context of $q-$orthogonal polynomials theory.

Although we didn't consider such possibilities here, for $q$ a root of unity, or dynamical integrable boundary conditions\,\footnote{For integrable dynamical boundary conditions, the corresponding open spin chain also enjoys a hidden $q-$Onsager dynamical symmetry \cite{dynam}.}, or the presence of inhomogeneities and higher spin representations located at each sites, it is clear that our approach can be applied analogously: in this case cyclic or higher tensor product finite dimensional representations of the tridiagonal algebra need to be considered in details (see e.g. \cite{qOns}). In this direction, infinite dimensional representations of the tridiagonal algebra also clearly need further investigations: they should play a crucial role in the analysis of the thermodynamical properties of the spin chain ($N\rightarrow\infty$) for generic boundary and anisotropy parameters, still unknown.

To conclude, using the complete basis of eigenstates (\ref{vectN}), (\ref{vecttrunc1}) or (\ref{vecttrunc2}) the problem of calculating correlation functions in the XXZ open spin chain with integrable non-diagonal boundary conditions (\ref{H}) can be handled. This will be the subject of a separate work.

\vspace{0.4cm}

\noindent{\bf Acknowledgements:} P.B. thanks N. Cramp\'e, O. Brodier and J. Garaud for help on MAPLE procedures. K.K. thanks CNRS and the French Ministry of Education and Research for financial support, and LMPT for hospitality. P.B wishes to thank the organizers of the 4-th Annual EUCLID Meeting 2006 where part of the results were presented. Part of this work is supported by the ANR research project ``{\it Boundary integrable models: algebraic structures and correlation functions}'', contract number JC05-52749 and TMR Network EUCLID ``{\it Integrable models and applications: from strings to condensed matter}'', contract number HPRN-CT-2002-00325.\vspace{0.5cm}

\vspace{0.7cm}

\centerline{\bf \large Appendix A: Coefficients of the block tridiagonal matrices}
\vspace{3mm}

All coefficients below have been calculated using the action of the nonlocal operators as defined in (\ref{rep}) on the family of states (\ref{eigenvectN}). For $l=0,1,...,N-1$ we find:\\

$\!\!\!\!\!\!\!\!\!${\bf Coefficients $\acal^{(N,l)}_{n[ij]}$:}\vspace{0.3cm}\\
For \quad $i,j\in\{1,...,\binomNmun\}$:\vspace{-0.3cm}\\
\beqa
&& \acal^{(N,l)}_{n[ij]}=  \frac{\cosh\phi}{\cosh^2(\phi/2)}\acal^{(N,l-1)}_{n[ij]} + \frac{\cosh\phi}{\cosh(\phi/2)}\dcal^{(N-1,l-1)}_{n[ij]} - \frac{1}{\cosh(\phi/2)}\acal^{(N-1,l-1)}_{n[ij]}\nonumber\\
&&\qquad \qquad+ \  \frac{\sinh(\alpha+(N-2-2n)\phi/2)}{\sinh(\alpha+(N-1-2n)\phi/2)}\left(\acal^{(N-1,l)}_{n[ij]}-\dcal^{(N-1,l-1)}_{n[ij]}\right) \nonumber\\
&&\qquad \qquad+ \ \frac{1}{2\cosh^2(\phi/2)\sinh(\alpha+(N-1-2n)\phi/2)}\hcal^{(N-1,l-1)}_{n[ij]}\!\!\!\! .\nonumber\\ \nonumber
\eeqa
For \quad $i\in\{\binomNmun+1,...,\binomN\}, \quad j\in\{1,...,\binomNmun\}$:\vspace{-0.3cm}\\
\beqa
&&\acal^{(N,l)}_{n[ij]}=  \frac{\cosh\phi}{\cosh^2(\phi/2)}\acal^{(N,l-1)}_{n[ij]}  -\frac{2e^{\alpha+(N-2n)\phi/2}\sinh(\phi/2)\cosh^2(\alpha+(N-2n)\phi/2)}{\cosh(\phi/2)\sinh(\alpha+(N+1-2n)\phi/2)}\ccal^{(N-1,l-1)}_{n[i-\binomNmun\ j]}\nonumber\\
&&\qquad \qquad+ \frac{e^{\alpha+(N-2n)\phi/2}\sinh(\phi)}{\sinh(\alpha+(N+1-2n)\phi/2)}\ccal^{(N-1,l)}_{n[i-\binomNmun\ j]}\ .\nonumber\\ \nonumber
\eeqa
For \quad $i\in\{1,...,\binomNmun\},\ j\in\{\binomNmun+1,...,\binomN\}$:\vspace{-0.3cm}\\
\beqa
&&\acal^{(N,l)}_{n[ij]}= \frac{\cosh\phi}{\cosh^2(\phi/2)}\acal^{(N,l-1)}_{n[ij]}    
+\frac{2e^{-\alpha-(N-2n)\phi/2}\sinh(\phi/2)\cosh^2(\alpha+(N-2n)\phi/2)}{\cosh(\phi/2)\sinh(\alpha+(N-1-2n)\phi/2)}\bcal^{(N-1,l-1)}_{n-1[i\ j-\binomNmun]}\nonumber\\
&&\qquad \qquad- \frac{e^{-\alpha-(N-2n)\phi/2}\sinh(\phi)}{\sinh(\alpha+(N-1-2n)\phi/2)}\bcal^{(N-1,l)}_{n-1[i\ j-\binomNmun]}\ .\nonumber \\ \nonumber
\eeqa
For \quad $i,j\in\{\binomNmun+1,...,\binomN\}$:\vspace{-0.3cm}\\
\beqa
&&\acal^{(N,l)}_{n[ij]}=\frac{\cosh\phi}{\cosh^2(\phi/2)}\acal^{(N,l-1)}_{n[ij]} + \frac{\cosh\phi}{\cosh(\phi/2)}\dcal^{(N-1,l-1)}_{n-1[i-\binomNmun\ j-\binomNmun]} - \frac{1}{\cosh(\phi/2)}\acal^{(N-1,l-1)}_{n-1[i-\binomNmun\ j-\binomNmun]}\nonumber\\
&&\qquad \qquad + \ \frac{\sinh(\alpha+(N+2-2n)\phi/2)}{\sinh(\alpha+(N+1-2n)\phi/2)}\left(\acal^{(N-1,l)}_{n-1[i-\binomNmun\ j-\binomNmun]}-\dcal^{(N-1,l-1)}_{n-1[i-\binomNmun\ j-\binomNmun]}\right)\nonumber\\
&& \qquad \qquad - \ \frac{1}{2\cosh^2(\phi/2)\sinh(\alpha+(N+1-2n)\phi/2)}\hcal^{(N-1,l-1)}_{n-1[i-\binomNmun\ j-\binomNmun]}\ .\nonumber\\
\eeqa
{\bf Coefficients $\bcal^{(N,l)}_{n[ij]}$}:\vspace{0.3cm}\\
For \quad $i\in\{1,..., \binomNmunpun\},\quad j\in\{1,...,\binomNmun\}$:\vspace{-0.3cm}\\
\beqa
&& \bcal^{(N,l)}_{n[ij]}=  \frac{\cosh\phi}{\cosh^2(\phi/2)}\bcal^{(N,l-1)}_{n[ij]} + e^{\phi/2}\left(\bcal^{(N-1,l)}_{n[ij]} - \bcal^{(N-1,l-1)}_{n[ij]}\right).\nonumber\\ \nonumber
\eeqa
For \quad $i\in\{\binomNmunpun+1,...,\binomNpun\}, \quad j\in\{1,...,\binomNmun\}$:\vspace{-0.3cm}\\
\beqa
&&\bcal^{(N,l)}_{n[ij]}=  \frac{\cosh\phi}{\cosh^2(\phi/2)}\bcal^{(N,l-1)}_{n[ij]}   +e^{\alpha+(N-1-2n)\phi/2}\frac{\sinh(\phi/2)}{\sinh(\alpha+(N-1-2n)\phi/2)}\left(\acal^{(N-1,l)}_{n[i-\binomNmunpun\ j]}-\dcal^{(N-1,l-1)}_{n[i-\binomNmunpun\ j]}\right)\nonumber\\
&& \qquad \qquad -\ e^{\alpha+(N-1-2n)\phi/2}\frac{1}{2\cosh^2(\phi/2)}\frac{\cosh(\alpha+(N-1-2n)\phi/2)}{\sinh(\alpha+(N-1-2n)\phi/2)}\hcal^{(N-1,l-1)}_{n[i-\binomNmunpun\ j]}
\ .\nonumber\\ \nonumber
\eeqa
For \quad $i\in\{1,...,\binomNmunpun\},\ j\in\{\binomNmun+1,...,\binomN\}$:\vspace{-0.3cm}\\
\beqa
&&\bcal^{(N,l)}_{n[ij]}= \frac{\cosh\phi}{\cosh^2(\phi/2)} \bcal^{(N,l-1)}_{n[ij]}\ .\nonumber\\ \nonumber
\eeqa
For \quad $i\in\{\binomNmunpun+1,...,\binomNpun\},\ j\in\{\binomNmun+1,...,\binomN\}$:\vspace{-0.3cm}\\
\beqa
&& \bcal^{(N,l)}_{n[ij]}=  \frac{\cosh\phi}{\cosh^2(\phi/2)}\bcal^{(N,l-1)}_{n[ij]} + e^{-\phi/2}\frac{\sinh(\alpha+(N+1-2n)\phi/2)}{\sinh(\alpha+(N-1-2n)\phi/2)} \left(\bcal^{(N-1,l)}_{n-1[i-\binomNmunpun\ j-\binomNmun]} - \bcal^{(N-1,l-1)}_{n-1[i-\binomNmunpun\ j-\binomNmun]}\right) \ .\nonumber\\ \nonumber
\eeqa
{\bf Coefficients $\ccal^{(N,l)}_{n[ij]}$}:\vspace{0.3cm}\\
For \quad $i\in\{1,...,\binomNmunmun\}, \quad j\in\{1,...,\binomNmun\}$:\vspace{-0.3cm}\\
\beqa
&& \ccal^{(N,l)}_{n[ij]}=  \frac{\cosh\phi}{\cosh^2(\phi/2)}\ccal^{(N,l-1)}_{n[ij]} + e^{-\phi/2}\frac{\sinh(\alpha+(N-1-2n)\phi/2)}{\sinh(\alpha+(N+1-2n)\phi/2)}\left( \ccal^{(N-1,l)}_{n[ij]} -\ccal^{(N-1,l-1)}_{n[ij]}\right) \ .\nonumber\\\nonumber
\eeqa
For \quad $i\in\{\binomNmunmun+1,...,\binomNnmun\},\ j\in\{1,...,\binomNmun\}$:\vspace{-0.3cm}\\
\beqa
&&\ccal^{(N,l)}_{n[ij]}=  \frac{\cosh\phi}{\cosh^2(\phi/2)}\ccal^{(N,l-1)}_{n[ij]}\ .\nonumber\\ \nonumber
\eeqa
For \quad $i\in\{1,...,\binomNmunmun\}, \quad j\in\{\binomNmun+1,...,\binomN\}$:\vspace{-0.3cm}\\
\beqa
&&\ccal^{(N,l)}_{n[ij]}=  \frac{\cosh\phi}{\cosh^2(\phi/2)}\ccal^{(N,l-1)}_{n[ij]}   -e^{-\alpha-(N+1-2n)\phi/2}\frac{\sinh(\phi/2)}{\sinh(\alpha+(N+1-2n)\phi/2)}\left(\acal^{(N-1,l)}_{n-1[ij-\binomNmun]}-\dcal^{(N-1,l-1)}_{n-1[ij-\binomNmun]}\right)\nonumber\\
&& \qquad\qquad +\ e^{-\alpha-(N+1-2n)\phi/2}\frac{1}{2\cosh^2(\phi/2)}\frac{\cosh(\alpha+(N+1-2n)\phi/2)}{\sinh(\alpha+(N+1-2n)\phi/2)}\hcal^{(N-1,l-1)}_{n-1[ij-\binomNmun]}
\ .\nonumber\\ \nonumber
\eeqa
For \quad $i\in\{\binomNmunmun+1,..., \binomNnmun\},\quad j\in\{\binomNmun+1,...,\binomN\}$:\vspace{-0.3cm}\\
\beqa
&& \ccal^{(N,l)}_{n[ij]}=  \frac{\cosh\phi}{\cosh^2(\phi/2)}\ccal^{(N,l-1)}_{n[ij]} + e^{\phi/2}\left(\ccal^{(N-1,l)}_{n-1[i-\binomNmunmun\ j-\binomNmun]} - \ccal^{(N-1,l-1)}_{n-1[i-\binomNmunmun\ j-\binomNmun]}\right)\ .\nonumber\\ \nonumber
\eeqa
{\bf Coefficients $\dcal^{(N,l)}_{n[ij]}$}:\vspace{0.3cm}\\
For \quad $i,j\in\{1,...,\binomNmun\}$:\vspace{-0.3cm}\\
\beqa
&& \dcal^{(N,l)}_{n[ij]}=  \frac{\cosh\phi}{\cosh^2(\phi/2)}\dcal^{(N,l-1)}_{n[ij]} + \frac{\sinh^2(\phi/2)}{\cosh(\phi/2)}\acal^{(N-1,l-1)}_{n[ij]} - \frac{1}{\cosh(\phi/2)}\dcal^{(N-1,l-1)}_{n[ij]}\nonumber\\
&&\qquad \qquad+ \  \cosh(\phi/2)\dcal^{(N-1,l)}_{n[ij]}\ - \ \frac{\sinh(\alpha+(N-1-2n)\phi/2)}{2\cosh^2(\phi/2)}\hcal^{(N-1,l-1)}_{n[ij]}\!\!\!\! .\nonumber\\ \nonumber
\eeqa
For \quad $i\in\{\binomNmun+1,...,\binomN\}, \quad j\in\{1,...,\binomNmun\}$:\vspace{-0.3cm}\\
\beqa
&&\dcal^{(N,l)}_{n[ij]}=  \frac{\cosh\phi}{\cosh^2(\phi/2)}\dcal^{(N,l-1)}_{n[ij]}   -e^{\alpha+(N-2n)\phi/2}\frac{\sinh(\phi)\sinh(\alpha+(N-1-2n)\phi/2)}{\cosh^2(\phi/2)}\ccal^{(N-1,l-1)}_{n[i-\binomNmun\ j]}\ .\nonumber\\ \nonumber
\eeqa
For \quad $i\in\{1,...,\binomNmun\},\ j\in\{\binomNmun+1,...,\binomN\}$:\vspace{-0.3cm}\\
\beqa
&&\dcal^{(N,l)}_{n[ij]}=  \frac{\cosh\phi}{\cosh^2(\phi/2)}\dcal^{(N,l-1)}_{n[ij]}   +e^{-\alpha-(N-2n)\phi/2}\frac{\sinh(\phi)\sinh(\alpha+(N+1-2n)\phi/2)}{\cosh^2(\phi/2)}\bcal^{(N-1,l-1)}_{n-1[i\ j-\binomNmun]}\ .\nonumber\\ \nonumber
\eeqa
For \quad $i,j\in\{\binomNmun+1,...,\binomN\}$:\vspace{-0.3cm}\\
\beqa
&& \dcal^{(N,l)}_{n[ij]}=  \frac{\cosh\phi}{\cosh^2(\phi/2)}\dcal^{(N,l-1)}_{n[ij]} + \frac{\sinh^2(\phi/2)}{\cosh(\phi/2)}\acal^{(N-1,l-1)}_{n-1[i-\binomNmun\ j-\binomNmun]} - \frac{1}{\cosh(\phi/2)}\dcal^{(N-1,l-1)}_{n-1[i-\binomNmun\ j-\binomNmun]}\nonumber\\
&&\qquad \qquad+ \  \cosh(\phi/2)\dcal^{(N-1,l)}_{n-1[i-\binomNmun\ j-\binomNmun]}\ + \ \frac{\sinh(\alpha+(N+1-2n)\phi/2)}{2\cosh^2(\phi/2)}\hcal^{(N-1,l-1)}_{n-1[i-\binomNmun\ j-\binomNmun]}\ .\nonumber\\ \nonumber
\eeqa
{\bf Coefficients $\hcal^{(N,l)}_{n[ij]}$}:\vspace{0.3cm}\\
For \quad $i,j\in\{1,...,\binomNmun\}$:\vspace{-0.3cm}\\
\beqa
&& \hcal^{(N,l)}_{n[ij]}=  \frac{\cosh\phi}{\cosh^2(\phi/2)}\hcal^{(N,l-1)}_{n[ij]} + \hcal^{(N-1,l)}_{n[ij]} - \hcal^{(N-1,l-1)}_{n[ij]}\nonumber\\
&&\qquad \qquad
- \frac{\sinh\phi\sinh(\phi/2)}{\sinh(\alpha+(N-1-2n)\phi/2)}\left( \acal^{(N-1,l)}_{n[ij]} - \acal^{(N-1,l-1)}_{n[ij]} +\dcal^{(N-1,l)}_{n[ij]} - \dcal^{(N-1,l-1)}_{n[ij]}  \right)
\ .\nonumber\\ \nonumber
\eeqa
For \quad $i\in\{\binomNmun+1,...,\binomN\}, \quad j\in\{1,...,\binomNmun\}$:\vspace{-0.3cm}\\
\beqa
&&\hcal^{(N,l)}_{n[ij]}=  \frac{\cosh\phi}{\cosh^2(\phi/2)}\hcal^{(N,l-1)}_{n[ij]}  \nonumber\\ &&\qquad\qquad +2e^{\alpha+(N-2n)\phi/2}\sinh\phi\sinh(\phi/2) \frac{\cosh(\alpha+(N-2n)\phi/2)}{\sinh(\alpha+(N+1-2n)\phi/2)}\left(\ccal^{(N-1,l)}_{n[i-\binomNmun\ j]}-\ccal^{(N-1,l-1)}_{n[i-\binomNmun\ j]}\right)\ .\nonumber\\ \nonumber
\eeqa
For \quad $i\in\{1,...,\binomNmun\},\ j\in\{\binomNmun+1,...,\binomN\}$:\vspace{-0.3cm}\\
\beqa
&&\hcal^{(N,l)}_{n[ij]}=  \frac{\cosh\phi}{\cosh^2(\phi/2)}\hcal^{(N,l-1)}_{n[ij]} \nonumber\\ &&\qquad\qquad -2e^{-\alpha-(N-2n)\phi/2}\sinh\phi\sinh(\phi/2) \frac{\cosh(\alpha+(N-2n)\phi/2)}{\sinh(\alpha+(N-1-2n)\phi/2)}\left(\bcal^{(N-1,l)}_{n-1[ij-\binomNmun]}-\bcal^{(N-1,l-1)}_{n-1[ij-\binomNmun]}\right) 
\ .\nonumber\\ \nonumber
\eeqa
For \quad $i,j\in\{\binomNmun+1,...,\binomN\}$:\vspace{-0.3cm}\\
\beqa
&& \hcal^{(N,l)}_{n[ij]}=   \frac{\cosh\phi}{\cosh^2(\phi/2)}\hcal^{(N,l-1)}_{n[ij]} + \hcal^{(N-1,l)}_{n-1[ij]} - \hcal^{(N-1,l-1)}_{n-1[ij]}\nonumber\\
&&\qquad \qquad
+ \frac{\sinh\phi\sinh(\phi/2)}{\sinh(\alpha+(N+1-2n)\phi/2)}\nonumber\\ &&\qquad\qquad \times\left( \acal^{(N-1,l)}_{n-1[i-\binomNmun\ j-\binomNmun]} - \acal^{(N-1,l-1)}_{n-1[i-\binomNmun\ j-\binomNmun]} +\dcal^{(N-1,l)}_{n-1[i-\binomNmun\ j-\binomNmun]} - \dcal^{(N-1,l-1)}_{n-1[i-\binomNmun\ j-\binomNmun]}  \right)
\ .\nonumber\\ \nonumber
\eeqa
The ''initial'' conditions for any $N$ are given by:
\beqa
&&\acal^{(N,-1)}_{n[ij]}=\bcal^{(N,-1)}_{n[ij]}=\ccal^{(N,-1)}_{n[ij]}=\dcal^{(N,-1)}_{n[ij]}=0\ ,\nonumber\\
&&\dcal^{(N,0)}_{n[ij]}=\lambda_n^{(N)}\delta_{ij}\ ,\nonumber\\
&&\hcal^{(N,0)}_{n[ij]}=(q^{1/2}-q^{-1/2})\lambda_n^{(N)}\acal^{(N,0)}_{n[ij]}\ ,\qquad \hcal^{(N,-1)}_{n[ij]}=-(q^{1/2}-q^{-1/2})\cosh^2(\phi/2)\delta_{ij}\nonumber
\eeqa
and 
\beqa
\acal^{(1,0)}_{0[11]}&=&\frac{\cosh\alpha^*\sinh(\alpha-\phi/2)-\sinh(\phi/2)}{\sinh\alpha},\qquad
\acal^{(1,0)}_{1[11]}=\frac{\cosh\alpha^*\sinh(\alpha+\phi/2)+\sinh(\phi/2)}{\sinh\alpha}, \nonumber\\
\bcal^{(1,0)}_{0[11]}&=&\frac{e^\alpha(\cosh\alpha+\cosh\alpha^*)\sinh(\phi/2)}{\sinh\alpha}, \qquad \quad
\ccal^{(1,0)}_{1[11]}=-\frac{e^{-\alpha}(\cosh\alpha+\cosh\alpha^*)\sinh(\phi/2)}{\sinh\alpha}\ . \nonumber
\eeqa

\vspace{2cm}

\centerline{\bf \large Appendix B: Numerical results for $N=3$ sites}
\vspace{3mm}

For small values of $N$, arbitrary or special values of the boundary parameters and $q$ generic in (\ref{H}) we have checked numerically (using MAPLE) the block tridiagonal structure (\ref{actionIn}), (\ref{actionIs}), (\ref{actionIsdual}), (\ref{actionIndual}) of all nonlocal commuting operators. We have also checked that the eigenvalues obtained by direct diagonalization of (\ref{Ifin}) with (\ref{rep}) coincide with the ones derived from (\ref{spectrumI}) . The same computations have been done for the Hamiltonian: it shows perfect agreement between the eigenvalues derived from (\ref{H}) or (\ref{E}). Being technically reduced to the calculation of the roots of a characteristic polynomial of degree $d=2^N$ ($d=\mbox{dim}({{\cal W}}_\pm)< 2^N$ or $d=\mbox{dim}({\overline{\cal W}}_\pm)< 2^N$ in the truncated (algebraic) sector associated with Bethe-type solutions) and eigenvectors of the block tridiagonal matrix ${\cal I}_1^{(N)}$ (or truncated ones ${\cal I}_1^{(N)}|_{trunc}$, ${\overline{\cal I}}_1^{(N)}|_{trunc}$), the method can be easily extended to higher values of $N$. 

Below we report the results for $N=3$ as an example. On the right hand side of the tables, for generic boundary conditions the energy levels have been computed replacing (\ref{spectrumI}) into (\ref{E}). For special boundary conditions, the eigenvalues corresponding to the Bethe eigenstates written in the form $\Psi^{(N,-)}_{\Lambda_1^{(N)}}$ are underlined. They have been computed by plugging (\ref{spectrumIm}) into (\ref{E}). The remaining eigenvalues associated with the Bethe eigenstates $\Psi^{(N,+)}_{\Lambda_1^{(N)}}$ have been similarly computed using the complementary subset (\ref{spectrumIp}). Note that for $p=(N-1)/2=1$, one of the solutions to (\ref{eigenvPm}) and (\ref{eigenvPm2}) yields the {\it ground state energy} as expected from the results in \cite{Nepo}. \\

\vspace{2mm}

$\bullet$ {\bf Arbitrary boundary parameters:} For $\alpha=1.3+0.7i$, \ $\phi=1.2i$, \ $\theta=0.1$, \ ${\tilde\theta}=0.21$, \ $\kappa=1.2+0.11i$:\vspace{7mm}

\centerline{Eigenvalues} \vspace{-3mm}

\medskip
\centerline{
\begin{tabular}[t]{|c|c|c|} \hline
         $H_{XXZ}$ (\ref{H}) & $H_{XXZ}$ (\ref{E}) \\ \hline 
  2.970995654 & 2.970995806 \\
  2.198656210 & 2.198656341\\
  1.602395044 & 1.602395091 \\
  0.946000118 & 0.946000104\\   
  -0.019184890 & -0.019184648 \\ 
  -0.055723904 & -0.055723820\\
  -3.180523377 & -3.180523355\\
  -4.462614856 & -4.462614866 \\ \hline
\end{tabular}}

\medskip
\noindent

\vspace{8mm}

\vspace{8mm}

$\bullet$ {\bf Special boundary parameters:} \quad For $\alpha\pm{\tilde\alpha} = \ \ -i({\tilde \theta}-\theta) - (N-2p-1)\varphi/2\qquad mod(2i\pi)$\ ,

\qquad \qquad\qquad\qquad  \qquad \qquad\qquad\qquad \qquad \quad  $\alpha=1.3+0.7i$, \ $\phi=1.2i$, \ $\theta=0.1$, \ ${\tilde\theta}=0.21$:

\vspace{7mm}

\centerline{Eigenvalues for $p=0$} \vspace{-3mm}

\medskip
\centerline{
\begin{tabular}[t]{|c|c|c|}\hline
         $H_{XXZ}$ (\ref{H}) & $H_{XXZ}$ (\ref{E}) \\ \hline 
3.618207615 & \underline{3.618207522}\\
2.779649839 & \underline{2.779650028}\\
1.727013687 & \underline{1.727013757}\\
0.001649161 & \underline{0.001648996}\\
-0.026778318 & -0.026778318\\
-0.198835568 & \underline{-0.198835520}\\
-2.826978951 & \underline{-2.826979214}\\
-5.073927468 & \underline{-5.073927520} \\ \hline
\end{tabular}}

\medskip
\noindent

\vspace{5mm}

\centerline{Eigenvalues for $p=1$} \vspace{-3mm}

\medskip
\centerline{
\begin{tabular}[t]{|c|c|c|} \hline
         $H_{XXZ}$ (\ref{H}) & $H_{XXZ}$ (\ref{E}) \\ \hline 
2.493378633 & 2.493378482\\
2.412226563 & \underline{2.412226647} \\
1.848074870 & 1.848075023\\
1.636051825 & \underline{1.636051957} \\
-0.270698292 & -0.270698432\\
-0.304708338 & \underline{-0.304708400} \\
-3.835675623 & -3.835675507\\
-3.978649644 & \underline{-3.978649645}\\ \hline
\end{tabular}}

\medskip
\noindent

\vspace{5mm}

\centerline{Eigenvalues for $p=2$} \vspace{-3mm}

\medskip
\centerline{
\begin{tabular}[t]{|c|c|c|} \hline
         $H_{XXZ}$ (\ref{H}) & $H_{XXZ}$ (\ref{E}) \\ \hline 
2.423171820 & 2.423171828\\
2.358362091 & \underline{2.358362193} \\
1.963670555 & 1.963670532\\
1.210097504 & 1.210097648\\
-0.100200908 & -0.100200806\\
-0.193998388 & -0.193998355\\
-3.440275504 & -3.440275543\\
-4.220827174 & -4.220827207\\ \hline
\end{tabular}}

\medskip
\noindent

\vspace{8mm}

\vspace{8mm}

$\bullet$ {\bf Special boundary parameters:} \quad For $\alpha\pm{\tilde\alpha} = \ \ i({\tilde \theta}-\theta) - (N-2p-1)\varphi/2\qquad mod(2i\pi)$\ ,

\qquad \qquad\qquad\qquad  \qquad \qquad\qquad\qquad \qquad \quad  $\alpha=1.3+0.7i$, \ $\phi=1.2i$, \ $\theta=0.1$, \ ${\tilde\theta}=0.21$:

\vspace{5mm}

\centerline{Eigenvalues for $p=0$} \vspace{-3mm}

\medskip
\centerline{
\begin{tabular}[t]{|c|c|c|} \hline
         $H_{XXZ}$ (\ref{H}) & $H_{XXZ}$ (\ref{E}) \\ \hline 
 4.906122870 & 4.906122692\\
 3.889167375 & 3.889167292\\
 2.586895038 & 2.586895406\\
 0.451087518 & 0.451087370\\
 -1.198292710 & -1.198292642\\
 -1.359411089 & \underline{-1.359411137} \\
-3.053875109 & -3.053875627\\
-6.221693900 & -6.221693900\\ \hline
\end{tabular}}

\medskip
\noindent

\vspace{5mm}

\centerline{Eigenvalues for $p=1$} \vspace{-3mm}

\medskip
\centerline{
\begin{tabular}[t]{|c|c|c|} \hline
         $H_{XXZ}$ (\ref{H}) & $H_{XXZ}$ (\ref{E}) \\ \hline 
2.388082920 & 2.388082936\\
 2.313332883 & \underline{2.313332774}  \\
 1.826977133 & 1.826977088\\
 1.655209880 &  \underline{1.655209967}  \\
 -0.212564780 & -0.212564681\\
-0.234717115 & \underline{-0.234717212} \\
-3.807430794 & -3.807430776\\
-3.928890122 & \underline{-3.928890149} \\ \hline
\end{tabular}}

\medskip
\noindent
                         
\vspace{10mm}

\centerline{Eigenvalues for $p=2$} \vspace{-3mm}

\medskip
\centerline{
\begin{tabular}[t]{|c|c|c|} \hline
         $H_{XXZ}$ (\ref{H}) & $H_{XXZ}$ (\ref{E}) \\ \hline            
2.520902297 & 2.520902258\\
2.455383157 & \underline{2.455383158}  \\
1.907373082 & \underline{1.907373054}  \\
1.121112406 & \underline{1.121112526}\\
-0.101546424 & \underline{-0.101546492}\\  
-0.226017760 & \underline{-0.226017706}\\
-3.360315576 & \underline{-3.360315615}\\
-4.316891175 & \underline{-4.316891206} \\ \hline
\end{tabular}}

\end{document}